\documentclass[acmsmall, screen]{acmart}

\AtBeginDocument{%
  \providecommand\BibTeX{{%
    \normalfont B\kern-0.5em{\scshape i\kern-0.25em b}\kern-0.8em\TeX}}}

\setcopyright{acmcopyright}
\copyrightyear{2023}
\acmYear{2023}
\acmDOI{XXXXXXX.XXXXXXX}

\acmConference[CSCW '23]{Minneapolis '23: The 26th ACM Conference On Computer-Supported Cooperative Work And Social Computing}{October 14-18, 2023}{Minneapolis, MN, USA}
\acmBooktitle{Proc. ACM Hum.-Comput. Interact. 7, CSCW2, Article xx (November 2023)}
\acmPrice{15.00}
\acmISBN{978-1-4503-XXXX-X/23/11}

\usepackage{tabularx}
\usepackage{enumitem}
\usepackage{multirow}
\usepackage{ragged2e}
\usepackage{textcomp}
\usepackage{siunitx}

\begin{document}

\title[Social Wormholes]{Social Wormholes: Exploring Preferences and Opportunities for Distributed and Physically-Grounded Social Connections}

\author{Joanne Leong}
\authornote{Three authors contributed equally to this research.}
\email{joaleong@media.mit.edu}
\affiliation{%
  \institution{MIT Media Lab}
  \city{Cambridge}
  \state{Massachusetts}
  \country{USA}
}

\author{Yuanyang Teng}
\authornotemark[1]
\email{yt2448@columbia.edu}
\affiliation{%
  \institution{Columbia University}
  \city{New York}
  \state{New York}
  \country{USA}
}

\author{Xingyu "Bruce" Liu}
\authornotemark[1]
\email{xingyuliu@g.ucla.edu}
\affiliation{%
  \institution{UCLA}
  \city{Los Angeles}
  \state{California}
  \country{USA}
}

\author{Hanseul Jun}
\affiliation{%
  \institution{Stanford University}
  \city{Stanford}
  \state{California}
  \country{USA}
}

\author{Sven Kratz}
\email{svenkratz@icloud.com}
\affiliation{%
  \institution{Independent}
  \city{Mercer Island}
  \state{Washington}
  \country{USA}
}

\author{Yu Jiang Tham}
\email{yoojbruin@gmail.com}
\affiliation{%
  \institution{Snap Inc.}
  \city{Seattle}
  \state{Washington}
  \country{USA}
}

\author{Andrés Monroy-Hernández}
\email{andresmh@cs.princeton.edu}
\affiliation{%
  \institution{Princeton University}
  \city{Princeton}
  \state{New Jersey}
  \country{USA}
}
\affiliation{%
  \institution{Snap Inc.}
  \city{Seattle}
  \state{Washington}
  \country{USA}
}

\author{Brian A. Smith}
\authornote{Co-Principal Investigators.}
\email{brian@cs.columbia.edu}
\affiliation{%
  \institution{Columbia University}
  \city{New York}
  \state{New York}
  \country{USA}
}
\affiliation{%
  \institution{Snap Inc.}
  \city{Santa Monica}
  \state{California}
  \country{USA}
}

\author{Rajan Vaish}
\authornotemark[2]
\email{xingyuliu@ucla.edu}
\affiliation{%
  \institution{Snap Inc.}
  \city{Santa Monica}
  \state{California}
  \country{USA}
}

% 


\renewcommand{\shortauthors}{Leong, Teng, and Liu, et al.}

% \newcommand{\bruce}[1]{{\color{purple}\bf{BL: #1}\normalfont}}
% \newcommand{\yy}[1]{{\color{teal}\bf{YY: #1}\normalfont}}
% \newcommand{\hanseul}[1]{{\color{blue}\bf{For HJ: #1}\normalfont}}
% \newcommand{\joanne}[1]{{\color{orange}\bf{JL: #1}\normalfont}}

% % \def \revision #1{{\textcolor{blue}{#1}}}
\def \revision #1{\textcolor{black}{#1}}
%TC:ignore
\begin{abstract}
Ubiquitous computing encapsulates the idea for technology to be interwoven into the fabric of everyday life. As computing blends into everyday physical artifacts, powerful opportunities open up for social connection. Prior connected media objects span a broad spectrum of design combinations. Such diversity suggests that people have varying needs and preferences for staying connected to one another. However, since these designs have largely been studied in isolation, we do not have a holistic understanding around how people would configure and behave within a ubiquitous social ecosystem of physically-grounded artifacts. In this paper, we create a technology probe called \textit{Social Wormholes}, that lets people configure their own home ecosystem of connected artifacts. Through a field study with 24 participants, we report on patterns of behaviors that emerged naturally in the context of their daily lives and shine a light on how ubiquitous computing could be leveraged for social computing.
\end{abstract}
%TC:endignore
%   <concept>
%       <concept_id>10003120.10003130.10003233</concept_id>
%       <concept_desc>Human-centered computing~Collaborative and social computing systems and tools</concept_desc>
%       <concept_significance>500</concept_significance>
%       </concept>
%  </ccs2012>

% \ccsdesc[500]{Human-centered computing~Collaborative and social computing systems and tools}
\begin{CCSXML}
<ccs2012>
   <concept>
       <concept_id>10003120.10003130.10003233</concept_id>
       <concept_desc>Human-centered computing~Collaborative and social computing systems and tools</concept_desc>
       <concept_significance>500</concept_significance>
       </concept>
   <concept>
       <concept_id>10003120.10003130.10003131.10003234</concept_id>
       <concept_desc>Human-centered computing~Social content sharing</concept_desc>
       <concept_significance>500</concept_significance>
       </concept>
 </ccs2012>
\end{CCSXML}

\ccsdesc[500]{Human-centered computing~Collaborative and social computing systems and tools}
\ccsdesc[300]{Human-centered computing~Social content sharing}

%\keywords{ambient displays, augmented reality, smart glasses, awareness, social connection}
\keywords{}

\received{15 January 2023}
\received[revised]{15 April 2023}
\received[accepted]{8 May 2023}
\maketitle

\section{Introduction}
%\yy{another pass: emphasize Ubicomp more??}

\begin{figure}
    \includegraphics[width=14.5cm]{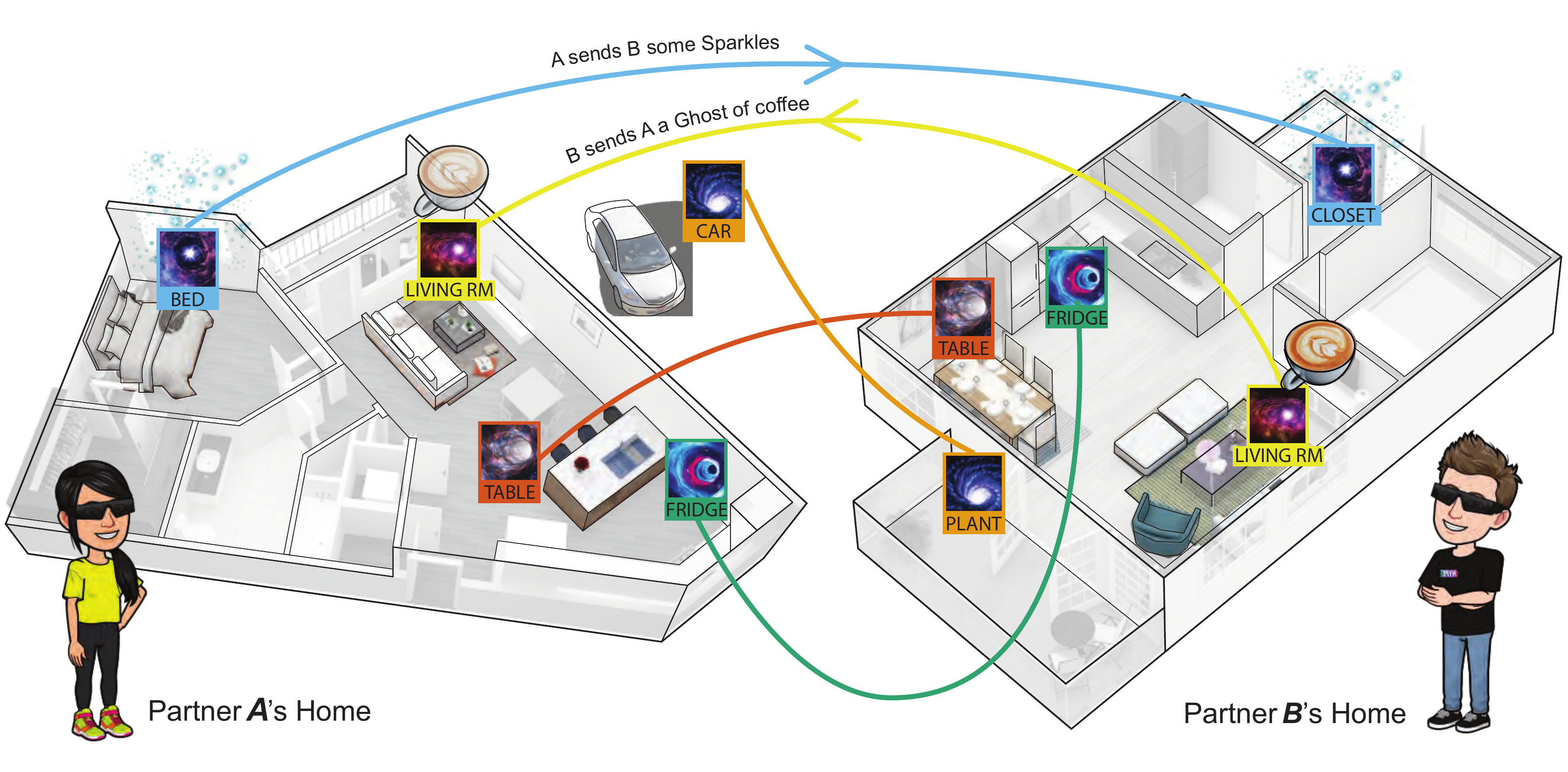}
    \caption{Illustration of the \textit{Social Wormholes} technology probe for fostering social connection, which uses physically-grounded connections and transmissions. The connections between two partners A and B are indicated by lines. The system supports symmetrical connections, such as table-table (red), fridge-fridge (green), and living room-living room (yellow). Asymmetrical connections are also supported, such as bed-closet (blue) and car-plant (orange). Partner A looks at her bed wormhole (left) through their AR glasses, which sends sparkles for Partner B to look at later when he visits his closet (right). Partner B captures the coffee mug in his living room (right) and transmits it as a Ghost to Partner A's living room (left).
    %Sparkles are sent from Partner a's bed wormhole to Partner b's closet wormhole. Ghost of a cup of coffee is sent from Partner b's living room wormhole to Partner a's living room wormhole.
    }
    \label{fig:teaser}
\end{figure}

Family and friends often desire to stay connected with each other over distance. Currently, our way of staying connected with others is device-centric, using smartphones, tablets, and computers. With the onset of smart homes, smart materials, and smart cities, however, the field of computing is slowly marching toward the vision of ubiquitous computing~\cite{weiser1991computer21century}, in which technology will evolve to become much more interwoven into our surrounding environment. 
A major opportunity for ubiquitous computing is to promote staying connected with others in an environment-centered way, repurposing the surrounding environment to serve as connection points with family and friends.
% physically grounded artifacts as means to connect family and friend. 
%putting family and friends in places and not just on devices. 
Very little work, however, has explored how ubiquitous computing and a potentially large ecosystem of connected endpoints might foster this sort of social connection.
% However, visionaries such as Mark Weiser have suggested that technology can evolve to be interwoven into the fabric of every day life \cite{weiser1991computer21century}. 
% However, pioneers such as Mark Weiser noted that the next frontier could be for technology to be interwoven into our everyday environments \cite{weiser1991computer21century}, such that the almost everything in a person's surrounding environment can be used as a medium for social connection.

A significant stream of work exploring social connection endpoints is that of \textit{Tangible Bits}~\cite{Ishii1997}, \textit{Ambient Media}, \textit{Ambient Telepresence}, and the idea that having connected objects or virtually connected spaces can be a powerful means of feeling close or even together with family and friends over distance. In these works, presence and activity information are embodied in physical artifacts including objects, surfaces, and spaces. For example, in AmbientROOM~\cite{ishii1998ambientroom}, information about a loved one's activities is relayed via the movement of a computer mouse and the projection of water ripples on the ceiling. Since then, numerous other systems have emerged that have been shown to be effective in supporting a sense of presence, awareness, and social connection in work and domestic life. These have taken on multiple forms, ranging from small household objects like candles, picture frames and desktop toys~\cite{chang2001lumitouch, brewer2005nimio, kim2015breathingframe, hakkila2018connectedcandles} to augmented furniture~\cite{dey2006awarenesstoconnectedness, siio2002peekadrawer}, and have evolved to relay various types of information, including mood, presence and activity between people~\cite{hassenzahl2012all}. 

% As such, these systems are fundamentally characterized by four design factors: what physical artifacts are used, how they are mapped together, what type of information can be transmitted between them, and the number of artifacts that make up the system. 
% knobs - justifies AR glasses 

% HOWEVER... 
% individuation \cite{ambe2017technologyindividuation} is a key aspect to creating augmented objects and that  
The vast array of design combinations is indicative that people may have a broad spectrum of needs and preferences for how to connect with one another via more ubiquitous social connections. Prior work echoes this sentiment, suggesting that ``one-size-fits-all solutions do not work'' for sharing aspects of domestic life~\cite{neustaedter2015sharingdomesticlife} and that people develop unique meanings when using augmented objects~\cite{ambe2017technologyindividuation, hassenzahl2012all}. What we notice, however, is that existing social connection systems are often designed for use in a fixed configuration, with few options for customization once they have been deployed. Many, for example, are based on a fixed pair of augmented objects with a dedicated set of specialized multi-modal capabilities. 
% \bruce{maybe be more specific, for example: they are designed to share one specific thing at one specific object or location} 
As such, user studies with these systems can reveal the impact those particular designs on social connection, but cannot lead to insights into user preferences beyond the scope of their specific design configurations. Additionally, while technology is maturing to a state where it might be possible to create a landscape of multiple connected artifacts, little is known about how users would perceive and use a broader ecosystem of distributed and physically grounded artifacts for social connection.
% \bruce{Probably need a core RQ here, just to start with: There is a lack of understanding about how people would want to share and maintain awareness with each other through object-based connections, if they were able to do so in an unconstrained fashion.}

Therefore, in this paper, we take a step back to explore what people's behaviors and resulting preferences would be if given the power to create their own ecosystems of connected objects for social connection. To do so, we create a customizable and scalable \textit{technology probe} \cite{Hutchinson2003} called \textbf{\textit{Social Wormholes}} and deploy it in a field study with 12 pairs of friends (24 participants) over two weeks. Figure~\ref{fig:teaser} illustrates \textit{Social Wormholes}. To make \textit{Social Wormholes} customizable and extensible for our investigation, we base its implementation on augmented reality (AR). With the platform, each person can make various physical artifacts in their homes points of social connection to their respective friend, by attaching printed AR markers. The markers are analogous to the concept of Tabs in Mark Weiser's original envisionment of ubiquitous computing~\cite{weiser1991computer21century}, turning any object into a social connection point quickly and cheaply. With the use of AR glasses and dedicated apps, connections can be established between the markers, and the artifacts can serve as mediums through which the friends send and receive snippets of information to each other. 

The flexibility of the probe makes it such that a pair of friends may (1) customize which objects in their homes become connected objects, (2) determine how these artifacts are mapped to one another, and (3) elect how many concurrent connections they use with their remote friend in the given period of time. Using this approach, we allow patterns of behavior to organically emerge, and capture this through a system of surveys, system logs and interviews. Based on an analysis, we report a spectrum of user preferences towards social connection. Furthermore, we outline a set of design recommendations for system features that may help to best serve these different populations. In summary, we contribute:

% we derive findings organically based on the collection and analysis of survey data, system logs, and interviews. From this, we report a spectrum of user preferences, and identify emergent patterns of behaviour amongst our participants. From this, we outline a set of design recommendations for features that may help to best serve these different populations. 
% when it comes to designing distributed and physically-grounded social connection systems. 

\begin{enumerate}
    % \item The design and implementation an AR-based technology probe that enables pairs of friends to establish multiple points of social connection with each other based on any physical object, surface, or place. From these points, they can communicate presence and other contextual information.
    \item The design and implementation of an AR-based technology probe of a distributed and physically-grounded social connection system for use between a pair of remote friends;
    % , based on a custom set physical artifacts (i.e. objects, surfaces or places)
    \item Insights from a field study with 12 pairs of friends (24 participants) on what types of physical artifacts people favor for social connection, how they could be mapped to one another, their perceptions around having a flexible number of connected artifacts, and different scenarios that can be supported with this infrastructure;
    \item An outline of different patterns of behavior users exhibit towards physically-grounded social connections, and associated design implications. 
    % scenarios and use cases emerged while using the system, and the types of information
    % \item A set of design implications given these groupings, and ideas for future research into distributed and physically grounded systems for social connection.
\end{enumerate}
\section{Related Work}
\label{sec:RelatedWork}
% Stemming from the idea that technology can advance to the point where all physical objects, surfaces and spaces can be connected,

There is a rich history of work on creating interconnected physical or digital objects for social connection, stemming from the visionary concept of Ubiquitous Computing \cite{weiser1991computer21century, weiser1997calmtechnology}. We provide an overview of prior works in this space that leverage physical artifacts for social connection, and also discuss frameworks for understanding how people design and interact with such technologies. 

%\subsection{Interconnected Artefacts for Interpersonal Connection}
\subsection{Ubiquitous Physical Artifacts for Social Connection}
% UbiMedia:: https://citeseerx.ist.psu.edu/viewdoc/download?doi=10.1.1.58.7776&rep=rep1&type=pdf

% Physical or Digital objects for social connection
% REPURPOSING EXISTING OBJECTS
% flexnfeel : https://dl.acm.org/doi/pdf/10.1145/2998181.2998247
% \yy{start with ubicomp and ubimedia as grandiose visions, then the subsequent rich history of connected artifacts are all attempts to realize that vision from one part or the other.}
% There is a rich history of developments in ambient media systems for social connection.
% Ambient media has traditionally spanned how to receive data that is both impersonal, such as stock prices, as well as information pertaining to people for improving social connection. However, we focus on systems that focus on fostering social connection. 
% There is a rich history of work on creating interconnected physical or digital objects for social connection, stemming by from the visionary concept of Ubituious Computing \cite{weiser1991computer21century, weiser1997calmtechnology}. 

Many prior works focus on improving social connection for people in Long Distance Relationships (LDR) or more generally, people who live in separate households. Examples of such systems include household fixtures and furniture such as bathroom mirrors, bed-side drawers, stools \cite{schmeer2010touchtracemirror, siio2002peekadrawer, dey2006awarenesstoconnectedness}, and household objects such as stuffed toys, candles, tools, picture frames and desktop toys \cite{fong2013bearwithme, hakkila2018connectedcandles, davis2019activityaallight, chang2001lumitouch, kim2015breathingframe}. Some works created entirely new physical artifacts such as novel desktop toys and radios \cite{brewer2005nimio, heshmat2020familystories} rather than augmenting existing ones. As pointed out by Li et al. \cite{li2018reviewemotionalcommunicationLDR}, a majority of systems rely on a single type of device for bidirectional communication, meaning they used a symmetric pairing of objects. There are also a few systems that feature non-matching connected objects. Examples of asymmetric systems, include Shared Wind \cite{yu2022sharedwind}, a uni-directional communication system with sender and receiver curtains, and Flex-N-Feel  \cite{singhal2017flex}, which comprised a flex-sensing sender glove and a vibrotactile receiver glove. A pioneering work, AmbientROOM, mapped \cite{ishii1998ambientroom} a pet's movement is mapped to a visual projection of ripples on the ceiling. 

User studies for these systems demonstrated that they were helpful in fostering improved feelings of presence and social connection. However, there is still much more room to investigate how people would use and be impacted by having ubiquitous socially-connected physical artifacts dispersed throughout one's environment. Therefore, with our technology probe, we enable users to designate existing artifacts in their homes as connection endpoints, establish how they are mapped to their remote friends' artifacts, and manage how many concurrent connections they use within their personalized ecosystem of connected artifacts. In this way, we can begin to gather insights into how people behave given the ability to create a ubiquitous constellation of social connections.

\subsection{Understanding Technology-Mediated Social Connection}

Many previous researchers have explored the design space for technology-mediated social connection. As Hassenzahl et al. \cite{hassenzahl2012all} explains, the feeling of ``relatedness'' is an integral psychological human need that technologies can support by supporting people in \textit{awareness}, \textit{expressivity}, \textit{physicalness}, \textit{gift giving}, \textit{joint action} and \textit{memories}.

% Investigations have been conducted from both a human needs and technical angle. These works do a lot to expand from the angle of mapping the design space and the various system variables and design axes.

% We sample from these papers what variables we need to address.
Given the wealth of strategies, many efforts have been made not only to create instances of these technological systems, but to understand and map out their design space. For instance, many researchers \cite{matthews2003peripheraldisplaytoolkit, pousman2006taxonomyofambientinfosystems,gooch2011designframeworkrelationshipdevices} have identified important design dimensions for ambient-media systems, ranging from information capacity and notification levels, to sensory mediums, personalization and more. In a systematic review of 150 articles of unconventional user interfaces for LDR emotional communication (i.e. excluding mobile apps) by Li et al. \cite{li2018reviewemotionalcommunicationLDR}, it was found that non-symmetric pairings of devices (meaning that the two objects are not of the same kind) and longer-duration studies of technologies for social connection in real-life use contexts remain underexplored and under-represented. Since we wish to expand knowledge in how a ubiquitous computing approach can be used for social connection, it was particularly necessary for us to incorporate these two aspects as part of our investigation. 
% For example, a designer could establish metaphors...

% Nevertheless, we explore patterns of behaviors, how users themselves interpret the meaning of the social wormholes they established.
% we're asking users, once they configured their environment,t any way they please, we ask them how they interpret the meaning of each endpoints that they setup. What are the user's own metaphors?
We base our investigative approach on a technology probe \cite{Hutchinson2003}, a research technique well-suited to studying social connection due to its ability to reveal surprising insights. For instance, Lottridge et al. \cite{lottridge2009sharingemptymoments} uncovered the power of promoting sharing during ``empty moments'' to nurture long-distance relationships. In another example, Judge et al. \cite{Judge2010,Judge2011} explore video-based platforms for social connection between remote families. They discovered that Family Window \cite{Judge2010}, which connected two households, triggered routine sharing of everyday moments, but that Family Portals \cite{Judge2011}, which connected three households, did not trigger such routine sharing. Grivas \cite{grivas2006digitalselves} experimented with establishing an imaginary ``merge'' of two homes using arrangements of physical LED prototypes in people's homes, and found that incorporating people's spatial knowledge of each other's places could evoke intimacy and a sense of presence. Unlike these previous technology probes, \textit{Social Wormholes} gives users a high degree of control over how to connect their space with their friend's space (users can connect anything with anything), allowing us to understand how people feel and behave with the ability to configure a ubiquitous computing environment specifically for social connection. 
% the ability to customi would choos ubiquitous system comprising a mosaic of different physically-based artifacts shapes feelings and behaviors around social connection. 
Our insights can inform the design of ubiquitous systems for social connection. 
% Grivas \cite{grivas2006digitalselves} began to explore this, by designing constellations of ``digital selves objects'' that transmit object usage information of ``twin objects'' (e.g. people's fridges) to facilitate intimate communications between homes. This work showed that topology and spatial interrelations between objects in the homes can contribute to a sense of presence and companionship.} 
% Grivas \cite{grivas2006digitalselves} experimented with establishing an imaginary merge of two homes using arrangements of physical LED prototypes in people's homes, and found that it could contribute towards a sense of presence of a remote person.}

%Although Wormholes employs AR to achieve its high degree of configurability, our focus in this work is not simply to create an AR-based awareness system but rather to investigate deeper questions around the idea of configurable awareness: namely, the use cases that emerge from such a system (RQ1), the patterns of connection that make sense for people (RQ2), and the values that abstract and concrete forms of awareness provide (RQ3). Prior works in AR-based communication explore the paradigm of ``leaving behind'' AR text~\cite{lin2011nunote, nassani2015tag, bace2016ubigaze} and audio messages~\cite{langlotz2013audio} at places so that others see them there later, as well as communicating with others via avatars in virtual spaces~\cite{choe2020augmented}. These works do not, however, explore the research questions that we explore here.

Other systems have explored AR as a means of sending and leaving physically-grounded messages for friends and colleagues. Specifically, they explore the paradigm of ``leaving behind'' AR text~\cite{lin2011nunote, nassani2015tag, bace2016ubigaze} and audio messages~\cite{langlotz2013audio} at places so that others see them there later. Unlike this body of work, our technology probe is centered around an ecosystem of connected physical artifacts rather than virtual AR objects being left behind. Other recent works in AR have explored using AR glasses for communication. For example, ARwand \cite{armessenger} enabled people to compose and send virtual content to be rendered on a remote friend's AR glasses, and for which they can see the friend's reactions. ARcall \cite{surale2022arcall} was a platform for a remote friend to ``drop in'' to see what a remote friend sees, and to inject an AR asset into their view to be seen via AR glasses. In contrast to these works, our AR-glasses-based technology probe does not involve direct sending of content to a person wearing the glasses, but rather allows people to transmit content anchored to virtually connected physical objects. The probe employs the use of AR to make the marginal cost of establishing new connected artifacts small---just more AR markers (pieces of paper)---giving our participants a chance to live in a large ecosystem of connected objects.

\section{Technology Probe: \textit{Social Wormholes}}

\begin{figure*}[b]
    \centering
    \includegraphics[width=10cm]{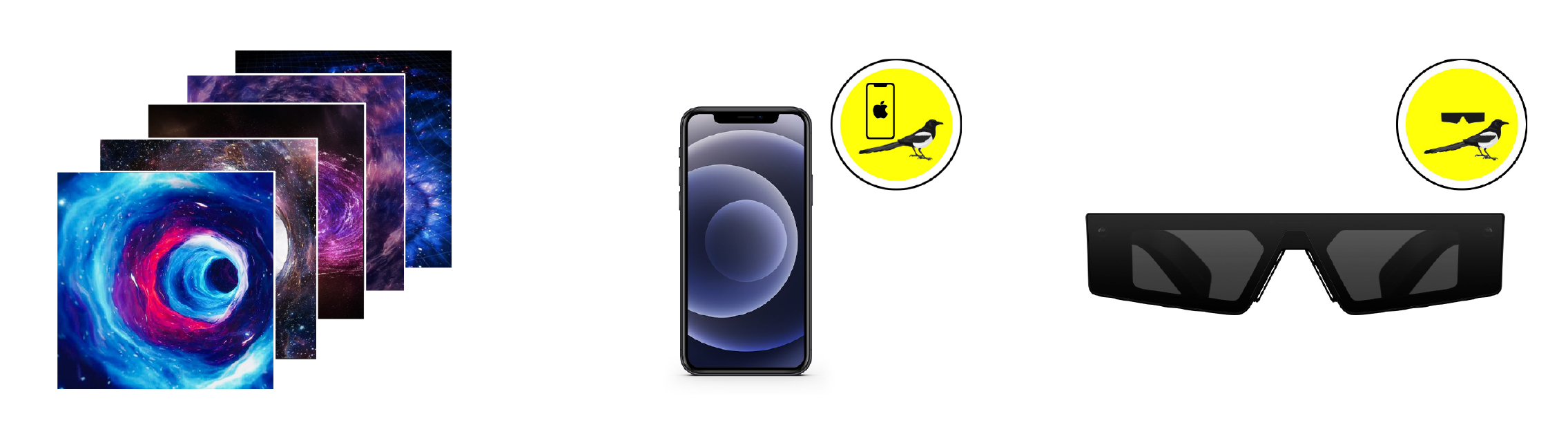}
    \caption{\textit{Social Wormholes} comprises three components (left to right): a set of printed markers which serve as endpoints for connections, a setup app for establishing connections between endpoints, and a connection app to receive and transmit content. Each person in a pair must have access to all components to use the system.}
    \label{fig:SystemComponents}
\end{figure*}

\textit{Social Wormholes} is an AR-glasses based system that serves as a flexible platform for social connection between two people. It has three components as seen in Figure \ref{fig:SystemComponents}: (1) a set of printed markers (i.e. wormhole images) that users can attach to objects or locations of their choice and distribute within their physical space; (2) a setup app for users to establish new wormhole connections or manage existing connections; and (3) the main AR-glasses application for users to send and receive Sparkles and Ghosts. 

    % Firstly, focusing on the ubiquitous computing perspective, it was important that the probe would enable a physically-grounded experience, meaning that...  meaning
    % the technology probe should be grounded to real physical artifacts in the home,
    % Thirdly, the probe should not enforce explicit and deliberate interactions, but rather enable users to engage in activities whilst also discovering or transmitting information. 

\revision{We arrived at \textit{Social Wormholes} through an iterative design process. Our research team considered many radically different designs for the technology probe, undergoing several rounds of brainstorming and creating several low-fidelity prototypes to develop our ideas.}

\revision{
At the start of the design process, we outlined three base requirements that the technology probe should fulfill in order for us to explore how a potentially large ecosystem of connected artifacts can support social connections. First, the technology probe should make it possible to convert any existing object or place into a medium for social connection. Second, the probe should scale to accommodate multiple artifacts while remaining cost-effective to deploy. Last, the probe should allow for communication ``on-the-go'', wherein a user can use the system impromptu in a way that does not impede their current activity. Users should be able to interact directly with connected artifacts and stumble upon transmitted messages naturally in their environment, rather than intentionally check a separate device.}
% % Last, the probe should allow for ambient interaction and afford spontaneity. 
% concurrent task?
% impromptu interaction -- spontaneous
% simultaneous 
% hands free
% person as message vs object as message? == environment-centered?
% choose to communicate 
% Last, the probe should allow for serendipitous interactions, wherein a user can engage in their normal activities and engage with the system impromptu, in a way that does not impede their .
% Last, the probe should allow for communication "on-the-go", wherein a user can engage in their normal activities and use the system impromptu, in a way that does not impede their current activity.
%Last, the probe should allow for communication ``on-the-go'', wherein a user can use the system impromptu in a way that does not impede their current activity.

 \revision{During the iterative process, we considered designs such as attachable hardware modules, a large network of touchscreen tablets that could be placed next to objects, and a smartphone app that utilizes AR to overlay virtual content onto objects. We converged on a system that combines AR glasses with paper markers because it satisfied our requirements best. The paper markers allow users to convert any object into a medium for social connection (akin to Mark Weiser's tabs~\cite{weiser1991computer21century}) and can be  scaled inexpensively to make more connected artifacts, while the AR glasses could be used to engage with the system while also carrying out other activities.
}

\subsection{Design and Implementation}
\begin{figure*}[]
    \centering
    \includegraphics[width = 1.0\textwidth]{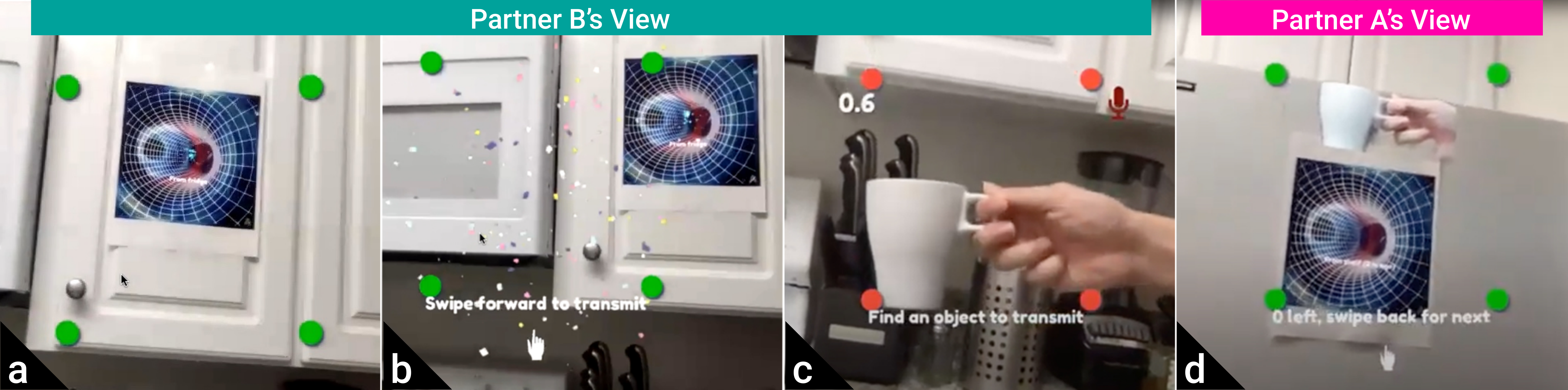}
    % \caption{Experiences as seen through AR glasses by connected partners A and B. From left to right: (a) B gazes at one of his Social Wormhole endpoints, which triggers Sparkles to be sent to A. (2) B views Sparkles at his shelf wormhole endpoint, which indicate that A had previously gazed at her connected endpoint. (3) B performs a swipe-forward gesture on the AR glasses that initiates a 5 second countdown during which audio is recorded. At the end of the countdown, the mug in his field of view is captured and transmitted along with the recorded audio as a Ghost to A. (4) Later, A receives the Ghost from B at the Wormhole endpoint on her Fridge.}
     \caption{\textbf{Experiences as seen through AR glasses by connected partners A and B.} From left to right: (a) B's AR glasses successfully detect the wormhole in his view, as indicated by the four green dots. Upon detection, B sees light blue Sparkles being emitted from his shelf wormhole endpoint, which indicates that A had previously gazed at her corresponding connected endpoint. (b) B's gaze towards his wormhole triggers Sparkles to be sent to A. (c) B decides to send A a Ghost, performing a swipe-forward gesture to initiate the process. A five-second countdown starts, during which audio for the Ghost is recorded. B holds a mug up in front of him, and it is captured at the end of the countdown. The Ghost, comprising both the mug and recorded audio, is then transmitted to A's corresponding wormhole. (d) Later, A receives the Ghost from B at the Wormhole endpoint on her fridge.}
    \label{fig:ARGlassesApp}
\end{figure*}

% \subsection{Social Wormholes System}
% Bruce;s Version
% because it is: (1) scalable, our system uses marker recognition and tracking to transmit and receive content between people, which means that one can establish any number of connections without specialized hardware installation that was required in traditional portal systems; (2) customizable, people can create and customize personal experiences by attaching wormhole endpoints to any objects and places; (3) integrated into routine, the effects applies anytime and anywhere without interrupting daily activities; and (4) private, only the wearer can see and receive information. 
We printed wormhole marker images at a size of 8$\times$8 inches on standard letter-sized sheets of paper. We tested the markers under various lighting conditions and viewing angles, and shipped five markers to each participant in order to ensure marker tracking accuracy and robustness. We developed two apps for our technology probe. First, we built a phone-based setup app (see Figure \ref{fig:PhoneApp}). Users can establish up to five one-to-one connections between wormholes of their choice. They can remove and change wormholes connections and setups by editing the connections in the app. In addition, we built a connection app on AR glasses (see Figure \ref{fig:ARGlassesApp}). With the connection app, users can transmit and receive content between the one-to-one wormholes connections they created. While other connection types may be valuable (i.e. one-to-many, many-to-many), we used this as a starting point to keep things simple for users (it is a common mapping pattern for prior systems, as noted in Section \ref{sec:RelatedWork}).
We developed both applications as Snap lenses using Lens Studio\footnote{https://lensstudio.snapchat.com/}. The \textit{Social Wormholes} system can operate on an iPhone 8 or later and on Snap Next Generation Spectacles. As part of the study, we lent AR glasses to the participants and confirmed that they would have access to a compatible smartphone. The setup app connects to the AR glasses via Bluetooth.

%\yy{perhaps add Information types here Sparkles and Ghosts}
Our system enables two formats of transmissions, \textit{Sparkles} and \textit{Ghosts}, as can be seen in Figure~\ref{fig:ARGlassesApp}. Sparkles are meant to be a lightweight form of communication. A user transmits Sparkles to a friend by simply looking at a wormhole while wearing the AR glasses. Once the gaze is registered, Sparkles are transmitted to the friend's corresponding wormhole. The next time the remote friend passes by their corresponding wormhole endpoint, they can receive the Sparkles by simply gazing at the wormhole marker while wearing their AR glasses. Once their gaze is recognized by their AR glasses, it plays an animation that shows the Sparkles in the form of particle bursts hovering above the wormhole opening on their end. Sparkles can act as an indicator that their remote friend was near their corresponding wormhole endpoint of a particular wormhole connection. Ghosts are a higher-fidelity form of communication that sends not just a single bit of information (as Sparkles do) but some richer context about the user's surrounding environment as well. A Ghost consists of the image of an object in the user's field of view accompanied by a short five-second audio recording. To begin capturing a Ghost, the user must look at a wormhole and perform a single-finger swipe-forward gesture on the touch pad of the AR glasses. This starts a five-second countdown, during which the surrounding audio is recorded (e.g., ambient noises, speech). At the end of the five seconds, an object in the user's field of view is captured using the forward-facing camera of the AR glasses. The captured object along with the audio recording is then sent to the friend's corresponding wormhole endpoint. The Ghost (captured object + audio recording) will appear for the friend the next time they look at their corresponding wormhole. This sequence is pictured in Figure \ref{fig:ARGlassesApp}.

% \joanne{An iterative prototyping and testing approach was taken by the authors internally, before arriving at the final design configuration. For example, the four dots as well as the animated instructions and wormholes were introduced to improve the usability based on initial impressions of the probe's design, and various marker sizes to strike balance between ease of use and detection.}
% Auggie -- based on these findings, our research team underwent several rounds of brainstorming to generate ideas for an effortful communication system. This involved each team member individually presenting ideas, followed by group discussion and iteration, and finally narrowing down on specific concepts that we felt best aligned with the design insights from our preliminary study and prior work. We also built and user tested several low-fidelity prototypes to further develop our ideas. We describe the final system design in the following section.

%Our system does background extraction by examining the saliency of the pixels using a pre-trained Machine Learning algorithm, keeping those with high saliency, and replacing the other pixels with transparency.

In order to facilitate the exchange of information between two connected users in a pair, we also implemented two cloud databases on Amazon Web Services (AWS) to store and readout Sparkles and Ghosts contents that users create. Specifically, we used DynamoDB to store wormhole endpoints connections, users, Sparkles and Ghosts transmission data, and S3 to store content like Ghosts images and audio recording. Both our phone app and AR glasses app communicate with the two databases to establish connections and support information exchange. Therefore, users needed a secure WiFi connection to use the system.

\subsection{Setup of Wormhole endpoints}
\begin{figure*}[]
    \centering
    \includegraphics[width = 1.0\textwidth]{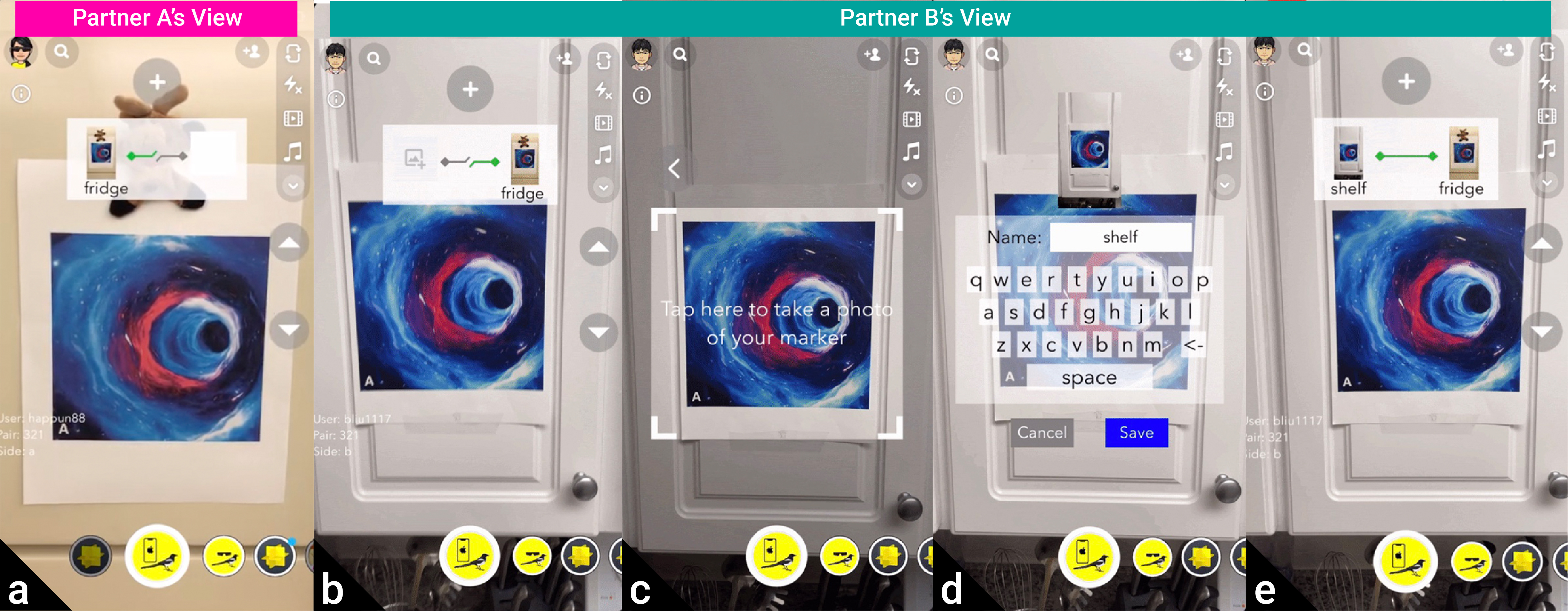}
    % \caption{Screenshots from the setup app. A pair of remote partners use the setup app to establish a Social Wormhole connection. Partner A makes the fridge her endpoint. Remote partner B completes the connection (indicated by the solid green line) by marking his kitchen shelf as his respective endpoint.
    % \label{fig:PhoneApp}}
    \caption{\textbf{Screenshots from the setup app} showing a pair of remote partners A and B use the setup app to establish a Social Wormhole connection. From left to right: (a) A initializes a Social Wormhole connection with her `fridge' as an endpoint. (b) Remote partner B views the incomplete connection initialized by A and taps the photo icon to begin to complete the connection. 
    (c) B taps to take a photo of his wormhole marker. (d) B names the marker `shelf' to correspond with its physical location. (e) This completes the connection, now indicated by the solid green line.
    \label{fig:PhoneApp}}
\end{figure*}

To explore the design aspect of location for a physically-grounded connection system, we designed \textit{Social Wormholes} to enable users to create an ecosystem of connected artifacts and configure the physical locations for their connection endpoints, as shown in Figure \ref{fig:teaser}. To create a wormhole connection, a user in a pair must first place the printed markers onto an artifact of their choice in their physical space. The user then initiates a connection in the setup app by taking a photo of their marker, and providing a text label to describe the connected artifact. Once saved, the incomplete connection will appear in a list to both partners in the app. The remote partner must then complete the connection in the setup app, by choosing the corresponding side of the incomplete connection, taking a photo of the marker on their chosen artifact and assigning it a label. In our technology probe, participants in each pair were given an opportunity to either freely place their markers, or coordinate with each other during the onboarding video conference call to set up their wormhole connections. Connected wormholes will appear in each user's list of connections with an image and label (Figure~\ref{fig:PhoneApp}). This process can be repeated for new wormhole connections. To edit the artifact of a wormhole endpoint, a user can simply move the printed marker to another artifact and update the label accordingly. 

% \joanne{
% % \subsection{Usability Testing}
% \subsection{Pretesting and Usability Iterations}
% We had team members who weren't involved with the implementation of the probe helped to test different design iterations. For instance, one of early design interations did not have the four dots in the field of view. However, we found through internal testing that people had difficulty ensuring that they kept their markers in the proper field of view of camera mounted on the AR glasses. Furthermore, we added animation effect an to show that it was active and to understand what was being transferred..
% }

\revision{
\subsection{Pretesting and Usability Iterations}
% During system implementation, system-developing researchers involved other researchers on the team to iteratively test the system. In the process, we discovered usability issues and iterated on the user interface design to improve the ease of use. For instance, an early implementation of the system did not provide visual reference to the AR glasses' camera field of view. From pretesting, we discovered that users had a hard time positioning and scanning the paper markers within the camera field view. We iterated on the design and added four dots denoting the boundary of the field of view. Furthermore, to clearly visualize that the camera is active and the messages are being transferred, we added animation effects during the iterative design process.
The team engaged in iterative prototyping and testing of the system. While a subset of the researchers focused on the system implementation, the others focused on testing it and providing feedback to improve the user experience. For instance, when pre-testing an early implementation of the system, we noted that some people had difficulty successfully scanning the paper markers with the AR glasses' forward-facing camera. We therefore iterated on the user interface design and added four dots to help people center the markers correctly within the camera's field of view for successful scanning. Our pre-testing also revealed that users sometimes struggled to understand when a transmission occured successfully. To improve this, we added animation effects to make this more apparent.
% The researchers who were not implementing the system tested it and provided feedback to improve the user experience.
}

\section{Methods}
In order to gain insight into how people would leverage and behave around a ubiquitous constellation of distributed and physically grounded artifacts for social connection, we designed and conducted a two-week field study with pairs of remote friends in their respective homes using the \textit{Social Wormholes} technology probe. \revision{Since this study was conducted within a technology company, it was reviewed and approved by the company's compliance, legal, and privacy teams prior to being conducted to ensure that the subject matter and approach complied with ethical standards and that participants’ data was processed appropriately.} 
%\revision{Our study proposal and protocol were reviewed internally by a panel consisting of independent privacy, compliance, and legal teams to ensure that the subject matter and approach complied with ethical standards and that participants’ data was processed appropriately.} 
In this section, we elaborate on the study procedure, measures, and approaches used for data collection and analysis. 
%\joanne{While we did not run a pilot for this study, the authors discussed and iterated on the design of the surveys and overall procedure to simplify it and maximize its clarity.}

\subsection{Participants}

\begin{table*}[t]
\caption{
    Self-reported demographic information of our participants who completed the study.
}
\label{tab:participants}
\tiny
\resizebox{\linewidth}{!}{
\begin{tabular}{lcclll}
\hline 
\textbf{Pair Number} & \textbf{Gender} & \textbf{Age} & \textbf{Education Level} & \textbf{Ethnicity} & \textbf{Relationship (IOS Scale \cite{aron1992iosscale})}
%& \textbf{Completed Study}
%1 = no overlap; 2 = little overlap; 3 = some overlap; 4 = equal overlap; 5 = strong overlap; 6 = very strong overlap; 7 = most overlap
\\ \hline
% \revision{P2a}  & F & 27 & Bachelor's Degree & East Asian & No\\
% \revision{P2b}  & F & 37  & Some College  & White & No \\
% \hline
P1a  & Man & 36 & Bachelor's Degree & White & 4 = equal overlap \\
P1b  & Man & 37 & Bachelor's Degree & South Asian & 3 = some overlap\\
\hline
P2a  & Woman & 30 & Bachelor's Degree & Not Reported & 2 = little overlap\\
P2b  & Woman & 27 & Master's Degree & East Asian & 2 = little overlap\\
\hline
P3a & Man & 28 & Master's Degree & East Asian & 5 = strong overlap\\
P3b & Woman & 27 & Bachelor's Degree & East Asian & 4 = equal overlap\\
\hline
P4a  & Woman & 23 & Bachelor's Degree & Hispanic & 4 = equal overlap\\
P4b  & Woman & 27 & Bachelor's Degree & White & 5 = strong overlap\\
\hline
P5a  & Woman & 21 & High School & East Asian & 1 = no overlap\\
P5b  & Woman & 21 & Bachelor's Degree & East Asian & 1 = no overlap\\
\hline
P6a  & Woman & 26 & Bachelor's Degree & South Asian & 1 = no overlap\\
P6b  & Woman & 30 & Master's Degree &  South Asian & 1 = no overlap \\
\hline
P7a & Man & 27 & Bachelor's Degree & White & 5 = strong overlap\\
P7b & Man & 27 & Master's Degree & White & 5 = strong overlap\\
\hline
P8a  & Man & 31 & Master's Degree & White & 7 = most overlap\\
P8b  & Man & 31 & Bachelor's Degree & White & 4 = equal overlap\\
\hline
P9a  & Man & 32 & Bachelor's Degree & White & 3 = some overlap\\
P9b  & Man & 30 & Some College & White & 3 = some overlap\\
\hline
P10a  & Man & 32 & Bachelor's Degree & White & 4 = equal overlap \\
P10b  & Man & 32 & Bachelor's Degree & White & 3 = some overlap \\
\hline
P11a  & Man & 29 & Master's Degree & Hispanic & 4 = equal overlap \\
P11b  & Woman & 30 & Bachelor's Degree & East Asian & 7 = most overlap \\
\hline
P12a  & Man & 30 & High School & Hispanic & 6 = very strong overlap \\
P12b  & Man & 28 & Some College & African American & 6 = very strong overlap \\
\hline

\hline
\end{tabular}}
\end{table*}

% We initially recruited 16 pairs of participants. 
%14 pairs of participants completed the on-boarding process. 
% Two pairs (Pair 1 and 6) were unable to complete the study due to COVID-related concerns. Two pairs (Pair 2 and 13) were unable to complete the study due to technical issues. In the end, 
12 pairs of participants (24 people total, including 10 females and 14 males) completed the study in full (see Table~\ref{tab:participants}) and were recruited from a technology company (name redacted) using an employee mailing list and internal communication channels. \revision{Since sample size for reaching data saturation cannot be predicted in advance, we determined our target sample size 10 $\pm$ 2 following recommendations by experts for empirical grounded procedures in HCI \cite{caine_samplesize}.}
% We eventually completed the study with 12 pairs of participants, meeting the target sample size.

\revision{The recruited participants were from a diverse set of teams within the company, representing a wide variety of technical and non-technical backgrounds including HR, sales/marketing and art/creativity. They were required to have access to WiFi in their homes and iOS smartphones with Bluetooth connection. Otherwise, there was no requirement for technical skills to participate in the study. Participants were located in 17 cities including Los Angeles, New York, London, San Francisco, Pittsburgh etc. Two pairs of participants, P9 and P11, were in different time zones from each other.} Participants were recruited in pairs who considered themselves friends, except for \revision{two pairs, P5 and P6,} that were formed ad-hoc and did not know each other before the study. Their ages ranged from 21 to 37 years (M = 29, SD = 4). Fifteen of them had a Bachelor's degree, six had a Master's degree, three completed some college, and two were high school graduates or equivalent. On a 5-point Likert scale (1 = Not at All, 5 = Extremely Frequently), participants on average reported not using AR/VR glasses at all (\(Mdn = 1)\). Each participant was compensated with a \$100 Amazon gift card for completing the study.
% We recruited from diverse departments within the company, representing a wide variety of technical and non-technical roles.
% We recruited from a diverse set of teams within the company, representing a wide variety of technical and non-technical background including HR, sales/marketing and art/creativity. Participants were located in 17 cities including Los Angeles, New York, London, San Francisco, Pittsburgh etc. Two pairs of participants (P9 and P11) were in different time zones from each other. 

% ages ranged from 21 to 37 years ($\mu$ = 29, $\sigma$ = 4)
%  participants reported 1.6 on average ($\sigma$=1.0)

% \subsection{Apparatus}
% For our study, we developed two applications: one for AR glasses and the other for the mobile phones.
% Both applications were developed using Lens Studio\footnote{https://lensstudio.snapchat.com/}. As part of the study, we lent AR glasses to the participants and confirmed about their access to an smartphone. 
%As the application for the mobile phones were for the setup stage required for using the AR headset application, every participant were given an AR headset and were confirmed their possession of a mobile phone.

\subsection{Procedure} 
% \subsubsection{Pilot Study}
% \subsubsection{Pre-study Onboarding}
% \subsubsection{During Study}
% \subsubsection{Post-Study}
% \yy{CHECK??}
% Before the study, each participant was given a set of five printed wormhole endpoints, and a pair of AR glasses and instructions for setting up the glasses. On the first day of participation, two researchers remotely on-boarded participants in pairs during one hour long sessions conducted over \revision{an interactive} video call. Researchers asked participants to fill out pre-survey questionnaires, helped participants set up systems required for the study, and thoroughly introduced them to the \textit{Social Wormholes} system.
% \revision{The researchers used interactive features via the video call to demonstrate and coach participants in operating the system, repeated as needed until participants were comfortable using it independently. After the on-boarding video call, we provided step-by-step visual instructions for the participants to reference during the field study. To support a smooth user experience throughout the field study, we dedicated an instant communication channel between each pair of participants and the researchers on Slack, to provide prompt assistance in case of any usability issues.}

Before the study, each participant was \revision{shipped} a set of five printed wormhole endpoints, a pair of AR glasses, and instructions for setting up the glasses. On the first day of participation, two researchers remotely on-boarded participants in pairs during one hour long sessions conducted over video call. Researchers asked participants to fill out pre-survey questionnaires, \revision{guided them in setting up their} systems, and \revision{and walked them through how to use} \textit{Social Wormholes}. \revision{The researchers answered all of the participants' questions to ensure that they understood the system's features and were comfortable using it independently. After the on-boarding session, participants received reference guides containing visual step-by-step explanations and instructions, which they could review when needed. Furthermore, the researchers created dedicated Slack channels for each participant pair. In this way, participants could contact the researchers for prompt assistance any time they had questions or concerns.}

Participants were asked to consistently use \textit{Social Wormholes} for at least 15 minutes per day for roughly two weeks. They were also asked to answer daily survey questionnaires that consisted of four open-ended questions about their usage through an online form. Participants used Wormholes for 11 days on average, with a maximum of 14 days. On the last day of the study, two researchers held an hour-long exit interview session with participants. Participants were interviewed in pairs. Finally, participants were also asked to complete an online exit survey. All surveys were issued via Google Forms. 

\revision{The studies were conducted in phases, to ensure that any issues (e.g. technical, usability, procedural, etc.) could be identified and addressed without affecting all samples. We planned to exclude participant pair data if they were to encounter any major issues. Initially, we deployed \textit{Social Wormholes} to two participant pairs in the first phase, followed by five pairs for each subsequent phase until we reached our target number of 12 pairs. Since no significant issues arose over the course of these phases, no data sets were discarded.}
% Throughout our phased deployments, we did not observe a need to discard our samples, since no significant usability issues arose.
% The studies were conducted in phases, to ensure that any usability issues could be identified and addressed
% allowing us to keep all of our samples.
% and there was no need to discard our initial samples.

% \revision{
% We anticipated that system and study variables might emerge during the field study, so we conducted our study in phases. We deployed the study to 2 pairs of participants in the initial phase. In case any variables emerge and require study design changes, this phase would have become a pilot study and we would have excluded the data. Fortunately, the initial phase went successfully and the study data met the \yy{standards?} to be included in our analysis. Subsequently, we deployed our field study in phases of 4 pairs of participants, until reaching a target sample size of 12 pairs. 
% }

% \revision{To ensure a smooth user experience, and to promptly respond to any usability issues, we dedicated a Slack channel, one for each pair of participants, and We 
% Interactive onboarding, visual reference, dedicated Slack channel one for each pair of participants tech support 
% }
% Responding to iterating on the study procedure 
% the first batch of people was really bad? then they become pilot... 
% we didn't run things in parallel , studies were not held concurrently, we were treating 
% in the end, we didn't have to make any significant changes 
% in the end, the data 
% first studies ironed out kinks 
% didn't catch any glaring errors 

\subsection{Measures}

\subsubsection{\textbf{App Log Data}}
We recorded log data of participants' usage of \textit{Social Wormholes} \revision{to gain insight on a more micro-level regarding how participants used the app's features}.
This includes information on markers --- when they were installed and what names were given to them, and content --- when were the ghosts and sparkles sent, which Wormhole endpoint they were sent from and sent to, and what content was transmitted. Data such as the Wormhole endpoint labels and the number of transmissions made per day were automatically logged by the phone-based setup app and the AR glasses connection app respectively. 
% \joanne{We used app log data to answer questions like what artifacts or places are being used for connection, and how do they engage with each other at these different locations?}
% \joanne{We utilized the log data in our qualitative and quantitative analysis to gain insight into how users generally engaged with the app, and what types of physical artifacts would be leveraged by them for social connection.}
%We utilized the log data in our qualitative and quantitative analysis. 
% we needed to understand on a micro-level how people were

\subsubsection{\textbf{Pre-Survey}}
\label{Sec:PresurveyQuestionnaire}
During the onboarding sessions, each participant completed a pre-survey \revision{so we could have insight into their backgrounds and the quality of their relationships prior to experiencing the technology probe}. This included questions on demographic information (i.e., age, gender, education level, ethnicity), previous AR/VR experience, level of loneliness using the short-form measure of loneliness (ULS-8) \cite{hays1987shortuls}, and the closeness of the partners' relationship with each other using the single-item Inclusion of Other in the Self Scale (IOS Scale) \cite{aron1992iosscale}. The ULS-8 \cite{hays1987shortuls}  contains 8 items (2 positively worded, which are reverse-coded), that should be rated on a 4-point scale (1 = Never 4 = Often). Examples of statements include ``\textit{I lack companionship},'' and ``\textit{I am an outgoing person}.''  The IOS Scale \cite{aron1992iosscale} asks people to select a single option on a 7-point scale (1 = Not at all close, and 7 = Extremely close), "Which picture best describes your relationship with your study partner?" A set of pictures with progressively overlapping circles, labeled ``Self'' and ``Other'' were provided to give a visual indication of the options available. See the supplemental materials for more details. 
% \joanne{These measures were collected to capture the participants' backgrounds and a pre-measure of the quality of their relationship prior to using the technology probe.}
% % The degree of closeness for the partners' relationship was measured using the single-item 
% The interpretation of the UCLA-20 is that the total score ranges from 20 to 80, with higher scores indicating greater loneliness. There is no specific cut-off score for saying whether someone is lonely or not. The score is more of a relative measure against the norm of the studied population. However, score of 20-40 are low to moderate, 40-60 is considered moderate to high, and above 60 is high. 

% level of loneliness using the short-form measure of loneliness (ULS-8) \cite{hays1987shortuls}, and the relationship with their study partner based on the Inclusion of Other in the Self Scale (IOS Scale) \cite{aron1992iosscale}.

\subsubsection{\textbf{Daily Survey}}
We asked participants to fill out a daily survey (refer to the supplemental materials for example questions) during their participation, which comprised multiple choice, Likert, and open-response questions \revision{in a Google Form}. \revision{We did this to capture the participants' thoughts and rationale behind their interactions over the course of using the technology probe.} Custom Likert scale questions based on specific features of the \textit{Social Wormholes} technology probe were issued, such as: ``\textit{Symmetric connections improved my sense of partner’s state over asymmetric connections},'' (1 = Strongly Disagree, 7 = Strongly Agree). They were also asked open-ended questions about what types of activities they engaged in over the day, whether they modified their Wormhole connections, what they tried to transmit, and how they perceived transmissions from their remote friend (e.g. ``What activities do you think your study partner engaged in while using \textit{Social Wormholes} today?''). Participants were reminded to complete the survey via daily calendar events that were scheduled in negotiation with them. %They completed them with respect to their prior full day using the probe. 
% \joanne{These questions were asked daily to capture the participant's thoughts and rationale behind their interactions as they used the technology probe.} 
% \revision{
%In order to safeguard the completion and quality of the daily survey, 
\revision{
% where participants must answer all questions in order to submit the survey. 
We periodically reviewed participants' responses to the daily survey over the course of the study to inform our exit interviews and to verify that participants were completing the questionnaire in good faith.}
% \revision{
% We were connected with participants daily on Slack channel, 
% }
% These questions were designed to capture their change in usage over the course of the field study. The survey also serves as a reminder for participants to take note of their experiences such that they can recount them during exit interviews. 
\revision{Given the time and effort involved in
filling out a survey every day, we required participants to fill out a minimum of one daily survey per week. }
% as well as messages on internal communication platforms
% The close-ended questions targeted specific type of connections and a role as a user--as a receiver and as a sender of given transmissions.
% The open-ended questions asked participants to report what they have sent and received, during what activity they have used the system, and what information they have learned about their partner.
% Rate your agreement with the statement

% we had an eye on their responses as they were trickling in 
% we kept in constant communication with them over slack
% we peeked at their responses as they came in, and we used that to create questions for their exit interveiw
% we kicked out participants... 

\subsubsection{\textbf{Exit Survey}}
At the conclusion of the field study, participants were asked to each answer an exit survey (see supplemental materials), \revision{so we could gain insight on how the technology probe impacted the quality of their relationship, as well as learn how their setups may have changed over time and why.} This exit questionnaire included questions that revisit their level of loneliness and relationship with their study partner using ULS-8 \cite{hays1987shortuls} and IOS scales \cite{aron1992iosscale} to capture a post-intervention measure (see descriptions of these scales in Section \ref{Sec:PresurveyQuestionnaire} or refer to the supplemental materials for more detail), and a longer and more comprehensive set of closed-ended and open-ended questions on their setup and usage of the technology probe. Examples of questions include, ``Where is Wormhole X, and how did you decide this?'', "X affected where I placed my Wormhole markers.'' One question also inquired about the sense of presence: ``While using \textit{Social Wormholes}, I can feel my study partner's presence in my space (1 = Strongly agree, 7 = Strongly disagree).'' 
% \joanne{These were taken as post-measures in order to gain insight on the impact the technology probe had on the quality of their relationship.}
% which was asked in the pre-survey questionnaire

% in a manner more comprehensive than the daily questionnaires.

\subsubsection{\textbf{Exit Interview}}
To conclude the study, the researchers conducted semi-structured interviews with participant pairs over video conferencing calls \revision{to elicit rich insights into their thoughts, feelings, and perspectives regarding their engagement with the probe}. Interviews were based loosely on the pre-prepared questions (see the supplemental materials). 
% that were used to elicit their thoughts and feelings with respect to using Social Wormholes. 
Unscripted follow-up questions were asked to dig deeper into points raised by participants. All interviews were recorded with participants’ consent and were later transcribed. Each interview lasted approximately 60 minutes. 
% \joanne{These were conducted in order to gain richer insights into the thoughts, feelings and perspectives participants had in engaging with the technology probe.}
%participants remotely met \revision{researchers} interviewing them over a video conferencing call.
% The interviews were conducted to gain more insights on the participants' \revision{perspectives, thoughts, and feelings with regard to using Social Wormholes.}
% \revision{Examples of interview questions are available in the supplemental materials.}
% and to gather information about their specific context of use and experiences.

% recent Critical Incident Technique (CIT) [ 36], in which we asked participants to recall and describe a recent time (or incident) where they explored an unfamiliar environment. 

% CIT is commonly employed by the HCI and CSCW community for conducting in-depth interviews [54, 60, 76, 111 ]. We adopted CIT because it gave us actual examples of times when our participants faced challenges in exploring unfamiliar environments, without having to undergo an extensive contextual inquiry and follow participants around all day long.
% However, as we discuss later in limitations (Section 7), future research could shadow VIPs to understand the context of their experiences in more detail and reveal more subtle nuances that may have been overlooked by using CIT

% behaviors, their comments and reflections around their experiences and preferences. 

%research themes we will introduce in the next Section, which are scenarios and use cases; placement and connections; and content/information type.

\begin{table}[b]
\caption{
Descriptive statistics for transmissions and connections, reported over the time duration of the entire study. Data in this table was collected from the setup and connection app usage logs. Total count includes everything created during the course of the field study across all participants.}
\label{tab:log}
\begin{tabular}{@{}lrrrr@{}}
\toprule
Asset            & Total Count & \%     & Mean/Person (SD) & \revision{Mean Count/Person/Day (SD)}\\ \midrule
% \revision{Endpoints (Pre)}  & \revision{129}            & \revision{}      & \revision{9.92 (0.65)}    \\
% \revision{Endpoints (Post)} & \revision{125}            & \revision{}      & \revision{9.62 (0.73)}    \\
\textbf{Transmissions}    & 2416           &        &                \\
\textit{Ghosts  }          & 576            & 23.8\% & 23.46 (14.01)  & \revision{1.89 (1.41)}\\
\textit{Sparkles}         & 1840           & 76.2\% & 75.33 (45.94)  & \revision{5.77 (2.95)}\\
\textbf{Connections }     & 54             &        &                \\
\textit{Asymmetrical }    & 39             & 72.2\% & 3.93 (1.22)    \\
\textit{Symmetrical}      & 15             & 27.8\% & 1.07 (1.22)    \\ \bottomrule

\end{tabular}
\end{table}

\subsection{Data Analysis}
\label{Sec:dataAnalysis}

\revision{We collected a} combination of quantitative and qualitative data using the aforementioned measures. \revision{We extracted t}otal counts for connections, endpoints, and transmissions from the app log data, and we further aggregated and analyzed \revision{these metrics per participant }pair (Table~\ref{tab:log}). Grand averages were computed from the IOS scale for relationship closeness, and from the ULS-8 scale for loneliness. Sense of presence were computed from the responses to survey questions (Table~\ref{tab:survey}).

% Quantiative Analysis
%  App log data was processed to derive quantitative metrics such as the total and average numbers of Wormhole endpoints created, the number of different locations and the number of transmissions. Data captured from the use of the ULS and IOS scales were processed to determine an average score per participant.

% Qualitative Analysis

% OLD 
% The exit interviews were transcribed and sectioned into quotes for a bottom-up, open-coding approach to data analysis.
% % transcripts were analyzed via a grounded theory approach ~\cite{charmaz_constructing_2006}. Following the grounded theory, w
% We iteratively coded the collected interview quotes, then merged their findings to identify the common themes. Two authors collaboratively performed open coding ~\cite{strauss_basics_1990} on the interview quotes to arrive at an initial set of codes. Then, affinity diagrams \cite{holtzblatt2017affinity} were built, and groupings emerged that highlighted an array of behaviors, perspectives, preferences, opportunities, and concerns. %to create constructive themes. 
% Several authors conducted weekly meetings to review and refine codes, themes, and design implications. We report on these themes in Section 6. 

% NEW
\revision{After transcribing the exit interviews, two researchers independently sectioned the transcripts into quotes for a bottom-up, open-coding approach to data analysis \cite{charmaz_constructing_2006}. Afterwards, the researchers worked together through multiple rounds of meetings to iterate on the codes, discuss their similarities and differences as part of a comparative analysis \cite{merriam2015qualitative}, and leverage them in an affinity diagramming process \cite{holtzblatt2017affinity}. The researchers determined that they reached code saturation when neither researcher could identify new codes or arrive at new interpretations of the existing codes after several rounds of revisiting the quotes. In accordance with Mcdonald et al. \cite{mcdonald_irr}, we did not compute inter-rater reliability (IRR), since we used the coding process to discover emergent themes or recurrent topics and permitted 
% and the researchers embraced multiple possible interpretatiosn of the meaing of the codes.
multiple possible interpretations of the meaning of the codes. 
After the two researchers completed their synthesis of an affinity diagram, two additional researchers reviewed the themes and provided their comments. The themes and sub-themes that emerged from this process highlighted an array of behaviors, perspectives, preferences, opportunities, and concerns, which we report on in Section 6.}

\section{Results: \textit{Social Wormholes} Usage}
24 participants (12 pairs) completed the study. 
% \revision{4 participants (2 pairs)} were dropped due to technical issues (e.g. failure to connect AR glasses to the internet) during onboarding, hence the smaller number of exit surveys compared to pre-\revision{surveys}. 
One person \revision{went on a trip} and had to re-establish the location of her wormholes during the course of the study. We logged 576 Ghosts and 1,840 Sparkles transmitted between participants, and collected 24 pre-survey responses, 118 daily survey responses, and 24 exit survey responses. 
% \joanne{CHECK: Participants used the probe on average for more than one session per day (M = 1.56).}
\revision{On average, participants used the probe more than one session per day (M = 1.56, SD = 0.32) at different times of the day.}
\revision{All participants completed the pre-survey, exit survey, and exit interview. All participants completed between two and ten daily surveys, on average more than four surveys, meeting our minimum requirement of at least one survey per week.}
% (M = 4.67, SD = 2.98)(with a mean of 4.92 daily surveys),

We report on our findings of how people used the technology probe during the study, including what artifacts people assigned as connections, how they chose to map them together, and how they used the flexible number of connected artifacts.

% General Usage
%\subsection{General Observations}
\subsection{General Usage and Self-reported Effects}
\label{SectionGeneralObservations}
%24 participants (12 pairs) completed the study. 1 pair (2 participants) was dropped due to technical issues (e.g. failure to connect Spectacles to the internet) during on-boarding sessions, hence the smaller number of exit surveys compared to pre-study questionnaires. One person traveled and had to re-establish the location of her Wormholes during the course of the study. 

\textit{Volume of Transmissions:}
Participants initially installed 129 wormholes with 47 types of artifacts (\textit{e.g.} desk, kitchen, wall, bedroom), and in their exit survey responses they reported 125 wormholes with 25 types of artifacts. During the field study, 1,840 Sparkles and 576 Ghosts were transmitted. Participants on average sent 23.46 Ghosts (SD = 14.01) and 75.33 Sparkles (SD = 45.94) to their study partners throughout the study. Refer to Table \ref{tab:log} for a summary of these metrics.

\textit{Effects on Social Connection:} 
% Participants reported that the technology probe significantly decreased their feeling of loneliness and increased the feeling of connectedness with their study partners. 
\revision{The pre- and exit-survey scores suggest that the technology probe generally decreased their feelings of loneliness and increased their feelings of connectedness with their study partners.}
The mean ULS-8 loneliness score was lower for the post-measure (M = 1.65, SD = 0.51) than the pre-survey measure (M = 1.77, SD = 0.49). \revision{While both means correspond with a generally low level of loneliness, 20 participants' loneliness scores decreased.} A one-sided Wilcoxon signed-rank test on participants' responses to the short-form measure of loneliness (ULS-8)~\cite{hays1987shortuls} before and after the technology probe (collected in pre-survey and exit survey respectively) shows a statistically significant decrease in participants' feeling of loneliness, with Wilcoxon statistic \(=144.5, p=0.022\). 
% \yy{In progress}\revision{The ULS-8 scale is generally used in a relative sense, on a range of 8-32, with higher numbers equating to higher levels of loneliness. Therefore we can say that the participants were not particularly lonely before and after the study. It decreased a little bit, with significance.}
Regarding relationship closeness, the median scores from the IOS Scale for the pre- and post-measures were equivalent (Mdn = 4). However Through an additional Wilcoxon test on participants' responses to the Inclusion of Other in the Self Scale (IOS Scale)~\cite{aron1992iosscale} before and after using Wormholes, we also observed a statistically significant increase in their connectedness ratings, with Wilcoxon statistic \(=5.5, p=0.0024\).  Statistical analysis shows that participants' responses indicated they were significantly less lonely and had stronger connections to their study partners after participating in our study and using Wormholes. Data on these metrics are included in Table \ref{tab:survey}.

\textit{Effects on Sense of Presence:} 
Alongside feelings of connection and loneliness, participants also indicated an increase in their sense of presence of their remote study partners in their physical space when using the technology probe. We observed a statistically significant correlation ($r=0.454, p=0.020$) between the number of transmissions (captured in the log data) and participants' ratings to the exit survey 7-point Likert scale question ``\textit{While using \textit{Social Wormholes}, I can feel my study partner's presence in my space.}'' (1 = Strongly Agree, 7 = Strongly Disagree). This suggests that participants who received more transmissions during the study reported a stronger sense of their study partner being with them. In the daily survey, P10b commented \textit{``I saw Sparkles from the desk area wormhole, which made me feel like they left their presence for me to discover.''} Data on this metric is included in Table \ref{tab:survey}.
% P9a\&b  described \revision{via the daily survey} that their communication through the connected artifacts brought a sense of ``\textit{being together}.''

\revision{
\textit{Impressions of AR and AR Glasses:}
As we describe in the beginning of Section 3, we chose an AR glasses-based approach for our technology probe because it gave us the properties that we needed to study how ubiquitous computing and a potentially large ecosystem of connected endpoints can foster social connection. The AR design itself was thus not a central focus of this investigation but more of a means to an end. Yet, because we used AR as a platform for this research, we were also able to glean insights about users' attitudes toward staying interconnected via AR and AR glasses.}

\revision{
 In exit interviews, several participants remarked that they appreciated how AR ``just makes [communication] more alive, rather than a flat picture or a flat video'' [P5a], compared to Snapchat, their usual means of staying connected with their study partner. This suggests that they saw value in rendering their communication itself in a more physical way, where the communication itself is manifested as a thing (an object).
 % This suggests that they saw AR as not just \emph{adding} matter (objects) to their environment but also as making their environment's existing matter smarter. That is, AR could enable existing objects to serve as vessels or mediums for maintaining connections with friends. 
 }

 \revision{
 Some participants felt that AR-based systems could lead to greater consequences for privacy since they require users to wear a camera or to point their phone's camera around their home. In the case of wearable cameras, these participants were unclear about what exactly was in the camera's field of view. We describe this and other privacy implications in Section~\ref{sec:privacy}.
}

%%%%%%%%%%%%%%%%%%%%%%%%%%%%%%%%%%%%%%%%

\begin{table}[]
\caption{
Statistics for survey questions.  Pre- and Post- measures were captured in the pre-survey and exit-survey respectively. A median is reported for relationship closeness and sense of presence since they were based on a single-item measure, while the mean is reported for loneliness, since the scale comprised multiple items.}
\label{tab:survey}
\begin{tabular}{@{}lrrrr@{}}
\toprule
                                                             & Pre      &      & Post       &      \\ \midrule
                                                             & Average  & SD   & Average    & SD   \\ \cmidrule(l){2-5} 
Relationship Closeness (IOS \cite{aron1992iosscale})                                 & Mdn = 4  & 1.86 & Mdn = 4    & 1.74 \\
\begin{tabular}[c]{@{}l@{}}Loneliness (ULS-8 \cite{hays1987shortuls})\end{tabular} & M = 1.77 & 0.49 & M = 1.65   & 0.51 \\
Sense of Presence                                            & N/A      & N/A  & Mdn = 3.96 & 1.67 \\ \bottomrule
\end{tabular}
\end{table}

% ``sense of your study partner being with you in your physical space''

% \begin{quote}
%     \textit{``There was a level of presence like `Oh, he's been at the fridge or he's been there' (P9a). Yes. He's been here (P9b). I'm here, we're both here (P9a).''}
% \end{quote}
%\yy{Discussion: This quote also reveals that participants interpret the connection as connecting spaces. }

% \joanne{WIP}
%  \joanne{We should see if there is a correlation between number of transmissions (not just number of Sparkles sent) and the "sense of your study partner being with you in your physical space'' ($r=0.454, p=0.020$)}
% From Sparkles, participants were able to indicate their partner's location.  `Oh, yes, [P4b] was here.''' -- P4a

% Limitation
% *** Due to the system design, sometimes things were shared unintentionally 

% We don't need to emphasize this anymore if we are not framing the system as an awareness specific system
% People can accurately understand the activities 

% Changing Wormhole Locations
% We might not need this information anymore. 
% Another aspect of our system is that it allows users to easily modify the locations of their wormholes. There were 105 wormholes whose initial and final locations during the study were understandable to us from participants' questionnaire responses.

% Point 1
\subsection{What Physical Artifacts Are Used?}
\label{sec:results-what-artifacts}
During onboarding for the technology probe, participants were guided to choose a vast array of physical artifacts in their homes to configure as wormhole endpoints, spanning from small objects, furniture pieces to wall surfaces of a room. With this customizability, we want to understand what types of physical artifacts people choose to use to stay connected and how they use them for communication.

\subsubsection{\textbf{What artifacts did people use?}}

% Between these, 40 wormholes did not change location during the study. Noticeably, there were 20 wormholes whose initial response was "desk," and 16 of them had the same response from the exit survey
% From our Wormholes system log data, wormholes spanned 47 artifacts but converged to a total of 25 by the end of the study. 
% \joanne{Do we have more info about where to where? As mentioned in the previous section, the locations of wormholes converged to popular places such as a desk or participants' fridge over time.}

%Participants established wormholes at different levels of representation of the space, from specific objects (coffee mug) to generic spaces (bedroom). 
%Participants configured wormholes using physical artifacts of different sizes, from small specific objects (coffee mug) to large generic spaces (bedroom). 
% \yy{This is repeating 5.1}
During set up based on log data, participants initially put their wormholes on 47 types of artifacts, with the top five most popular options being a desk (22), kitchen (11), fridge (6), wall (6), and door (6). In the exit survey, the participants were asked where the wormholes were by the end of the study and whether or not they had been moved. Wormholes converged to a smaller subset of 25 types of artifacts, the top six being a desk (29), kitchen (14), wall (9), bedroom (9), fridge (7), and door (7). Through the process of the study, participants have explored and thereafter discovered what artifacts they prefer to use as connections.
%\yy{why people moved some of the wormholes? is this a discovery process?}

This suggests that many participants chose to configure wormholes on common household objects and spaces. 
%like desks, kitchens and fridges, 
%\yy{add quotes for common household artifacts and spaces}
During exit interviews, some participants shared that they connected artifacts that are very personal and special rather than commonly intended for communications. 
P3b, P4a\&b and P10a centered their communication to their shared hobbies, such as playing piano [P3b, P10a] and sharing tarot cards. 
%P3a and P10a used their pianos as endpoints for communication:
% \begin{quote}
%     \textit{``I tried to show him my piano while singing to show him I'm a musical genius.''} - P10a
% \end{quote}

\begin{quote}
    \textit{``My favorite part was when I saw sparkles at his piano wormhole and heard him playing music --- it made me so happy as he has been helping me look for a piano to begin lessons and I remember him telling me that when he plays piano he feels happy or relaxed, so I was also happy for him!''} -- P3b
\end{quote}

P4a and P5b experimented with putting the wormhole on their pets or capturing playful moments of their pets\footnote{Participants confirmed that no animal was harmed during the study.}. 
%For example, P5b attached a wormhole on her cat and use it as a transmission endpoint to share cute moments of him:
\begin{quote}
     \textit{"I think it was pretty special when I sent you a ghost of my dog, which was also really hard to do because it kept moving around, and I just trying to track and follow them. "}  --P4a
\end{quote}

\begin{quote}
    \textit{``I just waited until he [P5b's cat] was laying down, and then the first time when I made the connection it worked, he was staying still and everything. Next time he was not friendly [laughter].''} -- P5b
\end{quote}

% \yy{How people consider artifacts as objects, spaces, people and activities}

\subsubsection{\textbf{How did people choose the artifacts?}}
\label{SectionHowDidPeopleChooseArtifacts}
% From the Likert-scale question in the exit survey, ``\_ affected where I placed my wormhole markers. (1 = Strongly Disagree to 7 = Strongly Agree),'' participants reported routine \revision{(\(Median=6, SD=0.7\)), environmental lighting (\(Median=6, SD=1.1\)), wormhole marker size (\(Median=4, SD=2.1\)), and household occupants (\(Median=4, SD=2.0\)) to be the factors that mattered the most for determining the placement of the wormholes.} 
From the Likert-scale question in the exit survey, ``\_ affected where I placed my wormhole markers. (1 = Strongly Disagree to 7 = Strongly Agree),'' participants reported routine (\(Mdn = 6\)), environmental lighting (\(Mdn = 6\)), wormhole marker size (\(Mdn = 4\)), and household occupants (\(Mdn = 4\)) to be the factors that mattered the most for determining the placement of the wormholes.

%Participants placed their wormholes in a manner that suited their daily routines rather than employing any preconceived spatial or social arrangement.
%a behavior aligning with our findings in Section~\ref{sec:use_cases}. 
Many participants placed their wormholes in a manner that suited their daily routines. This suggests that people desire connection technology to be seamlessly integrated into their existing environments, around spaces in their homes in which they regularly frequent while following their normal routines.
P5b, for instance, reported choosing artifacts specifically for connecting over everyday meals:
%P5b, for instance, reported placing his wormhole on his stove and using it to connect with his study partner everyday over meals:

\begin{quote}
    \textit{``It's by my stove, since I also visit the kitchen frequently. I thought it would be nice to connect with my partner over the food we are eating.''} -- P5b
\end{quote}
% \yy{Discussion: Connection as Connecting activity}

Artifacts that are located close to people's routines tended to get used more often than others. P4b explained that he used the wormhole in his kitchen the most because lots of routines happen around it:
\begin{quote}
    \textit{"It was just the one that I had the most to do around it. There were my coffee mugs there, there were my dishes, my food, my stove, so I thought about doing it more when I was at that one than any other one."} --P4b
\end{quote}

As another factor affecting where they placed wormholes, participants considered how well the wormholes fit into their space (echoing two of the factors found in exit survey: the wormhole marker size and other household occupants). P11 explained in the exit interview for not wanting the wormholes to clutter her space, while P3 elaborated that she wanted to customize the wormholes to fit her home decoration:

\begin{quote}
    \textit{``I just wish that instead of these markers, I had a cute little object that was unobtrusive and looked like a decoration that would trigger these things instead of a wormhole, because it doesn't go with the rest of my vibe in my house.''} -- P3b
\end{quote}

% \revision{P5 and P9 also mentioned that they tried to be mindful of spaces that they share with other members of their household, keeping wormholes in their private areas:}

% \begin{quote}
% \revision{
%     \textit{``Other people may not want to see markers in certain places. So [I] try to hide them.''} -- P9b
%     }
% \end{quote}

%yy{Another factor that points to how people choose what artifacts is special occasion. e.g. toilet for jokes, pool party}

%Additionally, one pair of participants [P7a \&b] commented that on occasions when they saw interesting things outside their residence, they wish there were a wormhole to share it. 
Additionally, one pair of participants [P7a \& b] commented that they would choose artifacts based on special occasions. When they saw interesting things outside their residence, they wished there was a wormhole to share it.
P7a brought the wormhole to places outside their routine so that they can capture interesting things there, such as pool and baseball games.
%One participant, P2a, tried to make jokes by placing the wormhole at unusual and awkward locations:
One participant, P2a, selected an unusual and awkward artifact for the sole purpose of making a joke:

\begin{quote}
     \textit{I also put it in funny places because I think it should be a light-hearted thing...I put it up with toilet where you sit on the toilet and it's on the wall right next to you. It's really funny.} - P2a
\end{quote}

%%%%%%%%%%%%%%%%%%%%%%%%%%%%%%%%%%%%%%%%%%%%%%%%%%%%
%%%%%%%%%%%%%%%%%%%%%%%%%%%%%%%%%%%%%%%%%%%%%%%%%%%%
%POINT 2

\subsection{How are Artifacts Connected to One Another?}
\label{SectionHowAreArtifactsConnected}

Social connection systems are often designed to be symmetrical, meaning that identical artifacts are used as a point of connection (e.g. A's candle is connected to B's candle). In this technology probe, participants chose to map things symmetrically and \textit{asymmetrically} (i.e. when the artifacts are different, such as a patio table to car). We report on how participants configured their connections to elucidate how people draw interpretations from social connections.  

% \subsubsection{Symmetric vs.\ Asymmetric Wormhole Connections}
\subsubsection{\textbf{Symmetric vs.\ Asymmetric Connections}}

Participants established 54 wormholes connections (15  symmetrical, 39 asymmetrical) in total with our technology probe over the duration of the field study. These metrics are summarized in Table \ref{tab:log}.
%Participants felt that symmetric connections improved awareness and made it easier for them to both understand their partner's state and share their own state. 
Participants indicated in the exit survey that symmetric connections improved the sense of presence with their partners, and made it easier to understand their partner's state and share their own state. Asymmetric connections led to special experiences.

In our daily surveys, participants agreed (rating of 4 or higher) with the statement \textit{``Symmetric connections improved my sense of partner’s state over asymmetric connections,''}  75.3\% of the time (\(Mdn=4\)); and agreed (rating of 4 or higher) with the statement \textit{``Symmetric connections make it easier to share my state over asymmetric connections,''} 80.4\% of the time (\(Mdn=5\)). For example, P9a\&b mentioned connecting their refrigerators to share cooking moments with each other, while P12a\&b described how symmetrical connections at their desks made their communication experience more immersive and helped them get a sense of working together. 
% \textit{``Symmetric connections improved my sense of partner’s state over asymmetric connections.''}  with a \textit{``4 -- Neutral'' or greater} 75.3\% of the time \revision{(\(Median=4, SD=1.8\))}; 
% and agreed with the statement \textit{``Symmetric connections make it easier to share my state over asymmetric connections.''} with a \textit{``4 -- Neutral'' or greater} 80.4\% of the time \revision{(\(Median=5, SD=1.6\))}.

% P12a stated that of their connections, he most frequently used their symmetric connection (desk-to-desk), and P12b felt similarly:

% \begin{quote}
% \revision{
%     \textit{``[P12b] had his desk as [wormhole] A, and I have mine at A as well. I feel like [P12b and I] have a lot more interaction in general [now], seeing what we're doing because [our wormholes] were in the same type of position. I can imagine himself typing when I'm typing.''} -- P12a
%     }
% \end{quote}
% \yy{Discussion: Connection as connecting the activity of typing together.}
 
Participants made many different types of asymmetrical connections including connections between one participant's TV [P2a] their partner's fridge [P2b], between one participant's plant [P2a] to their partner's toilet [P2b]. In addition, P2b kept all of their wormholes on their desk, while P2a scattered their wormholes throughout their house---their desk, TV, the wall next to their desk, their shower wall, and their car. While asymmetric connections might not be easier for participants to understand and share with each other, participants reported that the asymmetry produced a special, interesting experience:

\begin{quote}
    \textit{
    %``We did not have symmetrical connections, so I cannot speak to those. However, 
    "I do really enjoy asymmetrical experiences because my partner would have her wormhole on her cat for example, which will spice up the experience as a whole and keep it interesting.''} -- P5b
\end{quote}

%In summary, symmetric connections improved awareness while asymmetric connections led to special experiences.
% In summary, symmetric connections improved the sense of presence between partners, while asymmetric connections led to special experiences.

%\subsubsection{What physicially grounded connections mean to different users}
\subsubsection{\textbf{How Connected Artifacts are Interpreted by Users}}
%\yy{This result sections is too similar to Discussion 6.2.  Propose to only state findings such as "How users use and interpret the connections".  Leave "what they mean to users" to Discussion. }

We synthesize from daily surveys, exit surveys, and exit interviews that participants saw different degrees of value in the location of their artifacts. Some appeared to have a more spatial understanding of their connections; others focused on connecting over shared activities.
% others seemed to focus on connecting with people and sharing over activities.

Participants with strong spatial interpretation likened the experience to having a sense of presence [P10b] or leaving notes for each other around the house [P6a]. 
In daily surveys, participants reported that knowing where the transmissions are coming from adds significant value in understanding their partners' activities and daily routines. 
%P4a and P10b remarked
%, for example, that 
%they were able to infer rich information about what their partner was doing using Sparkles alone
%(which are very abstract) 
%simply because they knew where they were transmitted from.
%where the Sparkles had to have originated from. 

%repeated
% \begin{quote}
%     \textit{``I saw Sparkles coming from their door and their coffee machine, so I figured she must be starting her day; I saw lots of Sparkles by the kitchen and door, which made me think she was cooking before she left. ''} -- P4a
% \end{quote}

% \revision{quote removed}
% \begin{quote}
%     \textit{``I saw Sparkles from the desk area wormhole, which made me feel like they left their presence for me to discover.''} -- P10b
% \end{quote}

%likening it to leaving notes for each other around the house [P6a]. 
% However, P6a also described that having to receive transmission at specific locations creates a unique experience and a surprise feeling:
\begin{quote}
    \textit{``I would say it was very unique. %It's different from when you got a notification that you have a message or that you're getting a call because it's more passive. 
    It reminds me of my parents would leave notes around the house when they were not there. You're not expecting it and then you stumble on it.''} -- P6a
\end{quote}

%Others had a more abstract interpretation of the physically-grounded connections. 
Others, rather than placing significance on where their artifacts were located, simply saw the connections as a direct channel of information exchange to their partner. Participants rated the statement ``\textit{I felt that it was valuable to receive transmissions at particular places.}'' in  the exit survey with an overall neutral score (\(Mdn=4\)). 
For example, P11 stated that she prefers a direct 
%more traditional 
form of communication and would like the wormhole to be ``\textit{partner-based}'' instead of ``\textit{object-based}'':

\begin{quote}
    \textit{``It will be great if they can be customized, and I decide that in my room, this is [P11b]'s wormhole, and then I can glance at it any day I want, just to see her messages to me. ''} -- P11a
\end{quote}
% \revision{(\(Median=4, SD=1.5\))}. 

%\yy{Add examples where users interpreted the connections as doing some activity together }

Still others interpreted the wormholes connections as their partner's state of activity. 29 daily survey responses reported using wormholes to share activities. Our participants experimented with using Wormholes when working (reported 15 times in daily surveys), eating (14), doing chores (13), cooking (11), walking around (10), getting ready in the morning (8), watching TV (6), playing music (5), organizing/trying on clothes (5), driving (2), and doing laundry (1). 
In exit interviews, P10b stated that he frequently used wormholes to signify his work status. He tried to communicate a break in his work day - making coffee, singing to music, looking in the fridge. P2a reported working at his desk and transmitting a Ghost of his keyboard and mouse. P12a enjoyed connecting with their partner while working and experienced a strong sense of working together: 
% over food as well as other daily activities such as working. P12a commented on 
   \textit{"I can imagine himself typing when I'm typing"} -P12a . 
   %when he's sending me different types of stuff from the desk}. - P12a

Through a sequence of asynchronous transmitted artifacts, some users were able to interpret a sequence of transmissions as a continuous line of daily activities from their partner:
\begin{quote}
    \textit{``I saw Sparkles coming from their door and their coffee machine, so I figured she must be starting her day; I saw lots of Sparkles by the kitchen and door, which made me think she was cooking before she left. ''} -- P4a
\end{quote}

P7b and P5a stated they used the wormholes to track routine activities throughout the day, wherever they went:

% \begin{quote}
%     \textit{``Just a snapshot of whatever I was doing at the time. It was like, `Oh, I'll send what I'm eating in the morning,' and then I'll send like, `Oh, I'm at my desk working.' I'll send that. I put one on my door. When I'm going out somewhere, I'll send a Ghost from there.''}  -- P7b
% \end{quote}

\begin{quote}
    \textit{I'm at my desk, I'm drinking water. I said, I'm ghosting my water bottle or I just finished washing my face [...] I sent a my skincare or even my fan when it's hot outside of summer. [...] Oh yes, and then also washing the dishes.} -- P5a
\end{quote}

%For some people, they may consider where the other connected object is. For others, this is an unimportant detail, and rather, it is only important that friend can receive input regardless of their physical location.  
% Another is that an artifact simply is a channel to another person, and its precise location with respect to their friend can be ignored. 

% We observed that participants differed in their understanding of how they were connected through their artifacts. For some, they felt that artifacts would act as a direct channel to their partner. For others, they felt that the artifacts enabled them to be `together.'

%%%%%%%%%%%%%%%%%%%%%%%%%%%%%%%%%%%%%%%%%%%%%%%%%%%%
%%%%%%%%%%%%%%%%%%%%%%%%%%%%%%%%%%%%%%%%%%%%%%%%%%%%
% POINT 3? Below

% \subsubsection{Wormhole Endpoint Quantity and Distribution}
\subsection{How Many Artifacts Are Used?}
\label{SectionHowMany}
\textit{Social Wormholes} system log data and daily surveys show that 23 out of 24 participants who completed the study used more than 4 wormholes everyday. Most participants preferred to have large quantity of distributed connection endpoints.
%Participants who preferred a spread-out, large quantity of wormholes 
%They thought that 
In exit interviews, participants reflected that the distributedness provides better coverage and a more diverse communication experience with their friends:

\begin{quote}
    \textit{``[With] a greater number, I could definitely explore more options and get a more, a larger variety of contents. Just to learn more about like where she is at a moment versus having fewer. If we did have fewer wormholes, it might just be the same few things that are happening.''} -- P5b
\end{quote}

% how they configured which of their study partner's wormholes each of their own wormholes was connected to.

P5 and P6 also mentioned the distributed wormholes made the experience of maintaining awareness with their study partner feel like a ``scavenger hunt'' in a good way:

\begin{quote}
    \textit{``It was a treasure hunt of like, `Oh, where will I see it?' That experience was really cool.''} -- P6a
\end{quote}

Large number of wormholes also support spur-of-the-moment sharing and encourage participants to use wormholes for a variety of reasons. 
%We observed that they often use wormholes serendipitously, i.e., they were already next to the wormhole and made a spur-of-the-moment decision to share something very quickly.
31 comments from the daily surveys indicated that participants were motivated to use wormholes when they were physically around the wormhole location.
% Participants were motivated to use wormholes when they were physically around the wormhole location. This was reported 31 times in daily questionnaires.

On contrary, other participants mentioned in exit interviews that having a large number of wormholes might make it inconvenient for them to check messages.
Some preferred using only one wormhole or a centralized location for all wormholes. 
According to the log data, P12a mainly used one wormhole endpoint throughout the user study (i.e. used one of their endpoints more than 80 times, and used others less than 5 times). In the exit survey and interview, P11b shared how she compiled all wormholes together for ease of use:

\begin{quote}
\textit{`` I think in the beginning, we had [the wormholes] placed in different corners and I eventually just consolidated them. I had them in the pile near my desk. I always go through them in order.''} -- P11b
\end{quote}

\section{Results: Themes and Design Implications}

% \revision{We analyzed our findings via a grounded theory approach ~\cite{charmaz_constructing_2006}. Following the grounded theory, we iteratively refined our analytic frames and identified emerging themes. Two co-first authors collaboratively performed open coding ~\cite{strauss_basics_1990} on the interview transcripts to arrive at an initial set of codes. We synthesized the set of codes as necessary to arrive at theoretical saturation ~\cite{strauss_basics_1990}. Finally, several co-authors conducted weekly meetings to review the codes and refine them into a closed set of themes.}

%Here, we reflect on 
Following exit interviews open coding and affinity diagram analysis (see Section \ref{Sec:dataAnalysis}), we report on the set of themes of people’s behaviors and resulting preferences when given the power to create their own ecosystems of connected objects for social connection, which include people's broader patterns of behavior for using ubiquitous, physically-grounded social connections throughout their day (Section~\ref{sec:behaviorpatterns}) and what each of their connection endpoints represents to them (Section~\ref{sec:conceptualmodels}). We summarize our results and present our design implications for future physically distributed social systems in Table \ref{tabledesignimplications}. We also reveal how people adopt a mosaic of behaviors around how they use these social connections (Section~\ref{sec:mosaicofbehavior}), and other important social considerations for these social connections such as privacy (Section~\ref{sec:socialconsiderations}).

\begin{table}[t]
\renewcommand{\arraystretch}{1.2}
\caption{Summary of our study findings in terms of themes, insights, and design implications from Section~\ref{sec:behaviorpatterns} and Section~\ref{sec:conceptualmodels} for future physically distributed social connection systems.}
\label{tabledesignimplications}
\centering
\resizebox{\textwidth}{!}{

\begin{tabular}{{p{0.145\textwidth}p{0.41\textwidth}p{0.41\textwidth}}}
\toprule
 \hspace{0.4em}\textbf{Theme} & \textbf{Insight} & \textbf{Design Implications}\\
\midrule

\multicolumn{1}{l}{\multirow{2}{*}{\begin{tabular}{{p{0.145\textwidth}p{0.41\textwidth}p{0.41\textwidth}}}
% \revision{Ritualistic Communication}
Ritualistic \linebreak Communication
(Section \ref{SectionRitualistic})
\end{tabular}}}

& Users integrate physically-grounded communication in conjunction with daily routines and seek efficient and stable means of sending and receiving messages.
%Ritualistic behavioral pattern leads users to seek efficient and stable means of sending and receiving messages. Users often integrated physically-grounded communication in conjunction with daily routines.

& Keep records of the artifacts along the user’s routine, recognize changes in the user's routine over time and realign connected artifacts to their daily lives.
%The system should prompt the users to establish symmetry during setup and maintenance. 
%The system should 
Provide a centralized channel to recover messages for a sense of control and reassurance.
\\
\hline

\multicolumn{1}{l}{\multirow{2}{*}{\begin{tabular}{{p{0.145\textwidth}p{0.41\textwidth}p{0.41\textwidth}}}
Serendipitous Communication (Section \ref{SectionSerendipitous})
\end{tabular}}}

% & \revision{
%Serendipitous behavioral pattern motivates users to be experimental and creative about connecting with others.
% Users seek to be experimental and creative about connecting with others} 
& Users favor connections that are asymmetrical and changing because they add a sense of surprise to communication.
&

%Future systems should
Future systems can prompt users with creative ideas for connection points and mappings.
%Future systems could 
Expand connections to include one-to-many, many-to-one, and many-to-many connections.
\\
\hline

\multicolumn{1}{l}{\multirow{2}{*}{\begin{tabular}{{p{0.145\textwidth}p{0.41\textwidth}p{0.41\textwidth}}}
\RaggedRight Artifacts as Proxies for People \linebreak (Section \ref{Sec:ProxyforPeople})
\end{tabular}}}
& 
%Shrine Model
Some connected artifacts are interpreted as proxies for a person’s presence in ones space, similar to "shrines."
% \joanne{Users favor the use of artifacts that are specific reminders of their remote friends?}
&  

%Future systems could 
Future systems can keep records of the relationships between the user and their friends, offering suggestions for artifacts and spaces that suit the nature of the relationship. %(e.g., Desk for colleague, Coffee Table for friends).}
\\
% \hline
\cline{2-3}
& 
% wearable wormhole

%Users consider their 
Some connected artifacts are interpreted as direct channels to their partners, similar to a 1:1 chat. Users value instant updates from these, and they wish to carry the endpoints around as wearable devices.
& 

%Future systems could 
Sparkles and Ghosts can allow users to share their location in a less intrusive way than GPS or other means. Partners wishing to share their location in a low fidelity way may embrace the usage of Ghosts to give context.
% Design around the ability to get instant updates about their partner’s location. 
% Sender’s endpoints can be used as a lightweight form of location tracker and the receiver can be notified of their partner's location state instantly on the wearable wormhole.}
\\
\hline

\multicolumn{1}{l}{\multirow{2}{*}{\begin{tabular}{{p{0.145\textwidth}p{0.41\textwidth}p{0.41\textwidth}}}
\RaggedRight Artifacts as Overlapping \linebreak  or Connected \linebreak Spaces \linebreak (Section \ref{Sec:OverlappingSpaces})
\end{tabular}}}
& 
%Users consider the 
Some connected artifacts are interpreted as means of co-habiting spaces with partners for a sense of presence. Users prefer intuitive symmetrical mapping for this purpose.
& 

%Future systems could 
%Use computer vision to 
Recognize similarity between physical spaces and suggest symmetrical connections. Map the spatial co-relationship between the transmitted object and the wormhole and enable a representation of the transmitted object placed in the environment with spatial mirroring.
\\
\hline

\multicolumn{1}{l}{\multirow{2}{*}{\begin{tabular}{{p{0.145\textwidth}p{0.41\textwidth}p{0.41\textwidth}}}
\RaggedRight Artifacts as Shared Activities \linebreak(Section \ref{Sec:SharingActivity})
\end{tabular}}}
& 

%Users enjoy 
Some connected artifacts are used as channels for sharing particular activities. Users interpret interactions with the wormhole as  signals that the partner is doing the activity.
% interpret a stream of transmissions from their partner’s different wormholes as continuous daily activities. }
&  

%Future systems could 
Future systems can identify common activities that both connected partners engage in and suggest activity-signifying artifacts and spaces as connection points.
%a symmetrical mapping of connection points in the vicinity of that activity to the users. 
%Future systems could identify activity-signifying artifacts (e.g.piano) as its suggestions. 
% Last, some activities may benefit from one-to-many broadcasting or many-to-many connections. A multi-location alumni reunion is a good example of such an activity. Future systems could explore how to support such synchronous and scalable connections, and how to mitigate multiple sequences of activities in a network of physically-grounded connections.
% Support synchronous and scalable activities and mitigate multiple sequences of activities in a network of one-to-many or many-to-many connections.}
\\
\bottomrule
\end{tabular}}

\end{table}

\subsection{Behavior Patterns Around Communication}
\label{sec:behaviorpatterns}
%\subsection{Communicating Behavior Patterns}
%\subsection{Behavioral Patterns for Social Communication}
% How people communicate with the tech probe 
%\yy{rephrase "observation"}
% Based on our observation of user behaviors and comments 
Based on users' self-reports, descriptions, and comments of their usage from exit interviews, we discover two clusters of behavioral patterns toward exploring the technology probe as a communication medium: ritualistic and serendipitous. A ritualistic behavioral pattern leads users to seek efficient and stable means of sending and receiving messages. A serendipitous behavioral pattern motivates users to be experimental and creative about their process of connecting with others. 

\subsubsection{\textbf{Ritualistic Communication}}
\label{SectionRitualistic}
% Participants: 10b, 12 a \& b (excitement as receivers, want to know what they'll get) P9a\&b, P5a\&b
%We observed some users sought to establish a predictable and ritualistic pattern in their daily communication. They placed their Social Wormholes on artifacts they they'd use regularly, such as on their bedroom mirror to share their outfits in the morning, or the coffee machine to share about their breaks. 

Some participants [P4a, P5a\&b, P9a\&b, P10b, P12a\&b] establish a predictable and ritualistic pattern in their daily communication. They placed their wormholes on artifacts they were using everyday and integrated the physically-grounded communication in conjunction with routine moments of their life, 
%They ritualistically transmitted artifacts that represent their daily routine activities, 
such as placing one on their bedroom mirror to share their outfits in the morning or on their coffee machine to share information about their breaks. 

Both P4 and P9 reported sharing outfits of the day with their study partners habitually. P4a described when she was getting ready, she sent her outfit to her partner through the wormhole on the mirror, to show what she was wearing that day. P4b described her enjoyment receiving her study partner's daily update of her outfit:
    \textit{"I really enjoyed the Ghosts that she sent with her mirror because I could tell that she was having fun there.
    %I could tell that she liked that one. Why wouldn't you want it? 
    %That's most fun, to show your friend your outfit
    "} -- P4b
    
% \begin{quote}
%      \textit{For me, it was more of a daily thing. I would be making coffee. I did have one of the portals just close by so that I could use that on a daily basis.} - P12b
% \end{quote}

% \begin{quote}
%     \textit{I'm at my desk, I'm drinking water. I said, I'm ghosting my water bottle or I just finished washing my face [...] I sent a my skincare or even my fan when it's hot outside of summer. [...] Oh yes, and then also washing the dishes.} -P5a
% \end{quote}
Participants P5a and P7b described in detail how they shared a constant sequence of their routines throughout the day (Section 5.3.2). 
Participant pairs P5 and P9 developed a closer relationship with their partners by learning about their routine activities and environments:
% P9a \& b expressed that they gained insight into each other's daily life outside work, while P5a \& b felt they became closer friends as they learned more about each other's environments.
% learning more about their partner's routine patterns and environment. 
% For example, P9a \& b were coworkers at a tech company, and they learned about each other's daily life outside of work. P5a \&b were friends, they learned more about each other's environment and felt that their relationship was closer.

\begin{quote}
\textit{"He has a bit of like routine that I didn't know. I thought he just works and then goes and falls asleep until he comes back to work. I found out that he has a lot of the cooking routines and some of the stuff he would send me, just you can see a pattern."} - P9b
\end{quote}

% \begin{quote}
%     \textit{I definitely feel like I got closer to [P5a] through this activity. I learned that she had a cat [...] I got a little more about who [P5a] is and the environment that she's in on an everyday basis. I felt I really got to know her daily day-to-day life.} - P5b
% \end{quote}

In ritualistic communication, users' mappings were usually symmetrical, which made it intuitive for them to interpret the activity or space of their partners.
% They prefer to place the connection points on their daily routine, with a particular use case in mind for each connection point, such as sharing outfit when dressing up in front the mirror wormhole in the morning, then sharing coffee next to the coffee machine wormhole. The mapping of connections is usually symmetrical so that it is intuitive to interpret matching space or activity of their partners. 
To maintain a predictable pattern, they often desired to receive all the messages at expected places. 
This meant that they often only use a small set of fixed connections so that they would not miss out on any messages [P9b, P12b].
% These users were also reluctant to miss out on messages, and hence often used a smaller number of connection points [P9b and P12b].

\begin{quote}
    \textit{"For me, I would have preferred three or less. I think it would have just been more convenient and easier to keep up."} - P12b
\end{quote}

% People with an active lifestyle and a long routine curve around their house may need more wormholes, while people with a sedentary lifestyle favors to have a single wormhole near where they spent the most of their time, such as working desk or sofa. 

%They use only a limited set of connections such that they remember exactly where to look and .

% Both pairs P9a\&b and P5a\&b said during the user study they learned about each other's daily routine.

% \begin{quote}
%   Just a snapshot of whatever I was doing at the time. That's where originally, it was like, "Oh, I'll send what I'm eating in the morning," and then I'll send like, "Oh, I'm at my desk working." I'll send that. If I leave, I put one on my door. When I'm going out somewhere, I'll send a ghost from there. -P7b
% \end{quote}

Future systems can support a predictable and ritualistic communication experience by being designed to keep a long-term record to the connected artifacts along the user's routine. The system may also learn about the user's routine and recognize any small change of habits over time, in order to help users realign the connection points to their daily lives. Since symmetrical mapping is most intuitive for connecting partners, the system should prompt the users to establish symmetry during set up and maintenance. To avoid missing messages, users would appreciate a notification or a centralized channel to recover messages for a sense of control and reassurance. 

\subsubsection{\textbf{Serendipitous Communication}}
\label{SectionSerendipitous}
% (Note, to be removed: Participants: 10a, 11a, 3a, 4a\&b, 5a(occasionally), 7a\&b )

% Several participants emerge from study results display a serendipitous behavior pattern when they communicate with partners via the technology probe.
Several participants [P2a, P3a, P4a\&b, P7a\&b, P10a, P11a] exhibited a serendipitous tendency towards how they used \textit{Social Wormholes}.
Participants
%[P3b, P4a\&b, P5b, P7a\&b, P2a] 
also displayed serendipity by entertaining their partners with hobbies [P3b, P4a\&b], playing with pets[P4b, P5a], making jokes[P2a], or sharing interesting occasions and events[P7a\&b]. 
% When choosing what artifacts to connect and transmit, 
People behaving serendipitously found it important to incorporate some novel artifacts or something else that is different from the day before. They struggle to see the value of transmitting a routine object.

\begin{quote}
  \textit{
%   When I communicate with a friend like you want to communicate or maybe show something that you're doing, so it's more than an object. 
  "I found a hard time to everyday find one single object that I would want to send to [11b]... 
%   I don't know, 
maybe if I got a new pair of shoes and I want to send the one shoe to her."
%that's one example I can think of.
} - P11a
\end{quote}

% \begin{quote}
%     \textit{I would go out of my way to acquire more things to send, like I'll get a drink from the fridge or I'll turn the TV on, stuff like that just so I can change things up in there.} - P7a
% \end{quote}

%P3b, P4a\&b centered their communication to their shared hobbies, such as playing pianos and sharing tarot cards. P4b and P5a experimented with putting the wormhole on their pets or capturing ghosts of their pets. P7a \&b expressed the sentiment that when they saw interesting things outside their residence, they wish there were a wormhole to share it. P7a brings the wormhole to places outside their routine so that they can capture interesting things there, such as pool and baseball game. P2a tried to make jokes by placing the wormhole at unusual and awkward locations:

% \begin{quote}
%      \textit{I also put it in funny places because I think it should be a light-hearted thing...I put it up with toilet where you sit on the toilet and it's on the wall right next to you. It's really funny.} - P2a
% \end{quote}

% P3b has a wormhole at piano, and also wants to send tarot cards as a playful thing. 

% P4a and P4b shared music they listen to and music they play on piano to entertain each other. 

% Our findings revealed that, in order to explore a serendipitous experience in communication, users 
Users with serendipitous tendencies experimented with unconventional artifacts (i.e. non-standard household objects) as connection points. 
%The artifacts of choice are unconventional as they are beyond the common household physical locations and objects that most intuitive to use with the technology probe. The unconventional artifacts 
Examples include moving objects, pets, locations that provoke humor, and artifacts related to personal hobbies. The mapping of connections were often asymmetrical and changing, as the users preferred to spice things up, keep things fresh, and react to momentary changes of events. For both sending and receiving, users embraced a large number of connections so that they could share wherever and whatever they wanted and so that they could have the treasure hunt-like experience of finding messages by surprise [P6a]. 

Future systems can support and encourage serendipitous communication by adopting high customizability and flexibility in terms of what artifacts can be activated as connection points. The system should instigate the discovery of potential creative wormholes and transient changes in mapping connections. Changes in connected artifacts and mapping should be tracked and streamlined by the system rather than burdening users to provide accurate labels. 
% Because having a large number of connection points is advantageous for both sending and receiving,
Since a greater number of connection points leads to more spontaneous and surprising sending and receiving,
future systems could multiply the amount of surprising opportunities by expanding to include one-to-many, many-to-one and many-to-many connections. 

\subsection{What Physical Artifacts Represent}
\label{sec:conceptualmodels}
% How people formulate physically grounded connections % Connecting patterns ??
%Based on our observations of behaviors as well as their comments, participants used technology probe to configure physically-grounded connections with different perspectives in mind and interpreted their meanings differently. 
Learning from how artifacts are connected and interpreted by users, as well as users' comments and reflections, we notice that people configure physically-grounded connections with different ideas for what they represent. Artifacts could be interpreted in three ways: as a proxy for a person, as a means to connect a user's space with their partner's space, or as a means to share an activity with their partner. 

% \joanne{Can we highlight which participants had which attitude? It's very possible that participants showed a tendency to more than one of these.}

% In the first category, people leveraged the connected object as a direct channel to another person. This mirrors many previous works that focus on a connected pair of objects, where the object is dedicated to particular person. In this usage, the space in which the corresponding partner exists is not important/ 

% Ambient Media can give the sense that two people's spaces are connected or overlapping. Alternatively, a connected artifact can be interpreted as direct channel to someone -- independent from a physical position in space. 
\subsubsection{\textbf{Artifacts as Proxies for People}}
\label{SectionProxiesforPeople}
\label{Sec:ProxyforPeople}

% Participants: P4b, P7a \& 7b, 10b, 11a, 15b

% \joanne{sometimes they didn't even care that it piggy-backed on a real item...}

From their comments in exit interviews, we notice that several users [P2b, P4b, P7a \& 7b, P11a] considered their connected artifacts to be direct channels to---or proxies for---their partner. 
%For instance, P11a explicitly stated that they would like to dedicate a Wormhole to P11b: 

%repeated
% \begin{quote}
%     \textit{    ``...that'll be great if they can be customized, and I decide that in my room, this is my [11b] wormhole, and then I can glance at it any day I want, just to see her messages to me.''} -P11a
% \end{quote}
This behavioral pattern parallels the concept of reaching someone with a smartphone or landline phone. In this paradigm, the location of a partner's phone is not necessarily important; rather, the fact that a person can be reached through it as a medium takes precedence. 
Similarly, artifacts could serve as a proxy for a person by substituting for another person's presence in a space. P4b and their partner P4a both lived alone, but they interpreted the wormholes as representations of their partner in their respective spaces, like \textit{"having a low-impact roommate" }- P4b. 
While some participants stated one proxy was enough [P7a, P9b], others accepted the idea of having many proxies spread out across multiple locations in their homes. In this way, they could even provide a tour of their house [P10a].
% \textit{``I showed [P10b] different areas around my room throughout the day. I tried to show him my piano while singing to show him I'm a musical genius. My chopstick collection to show off my lips are on fleek, I made a joke that my closet was Narnia and I was looking for creatures in there.''} -- P10a. 
In yet another case, some people began to consider that even the content that was transmitted (and not just the wormhole itself) were proxies for their partners. For example, P7a and P7b came to associate the transmitted ghosts of their favorite beverages as stand-ins for each other.

We noticed that this general viewpoint for wormholes served two purposes: to act as "shrines" and to act as wearable connections. 

Regarding "shrines", people treating wormholes as proxies for other people found it important to mark a personal artifact of significance as a connection endpoint (e.g. a special figurine that signifies their friend) [P3b]. This mirrors former designs in Ambient Media that have a sentimental or personal nature ~\cite{chang2001lumitouch, heshmat2020familystories}.  
% (e.g. those based on picture frames where a loved one's photo gives the object a personal significance, such as with Lumitouch \cite{chang2001lumitouch}, or how connection devices were collectively placed next to personal gifts and postcards of a person \cite{heshmat2020familystories}).
For receivers, the location to where the physically grounded receiving endpoints are placed in their own environment is rather significant.
%value the spatiality of the artifact proxy in relation to their own space. 
%Although, for the receiver, it is not important to know the location of the sender, the location to where the physically grounded receiving endpoints are placed in their own environment is rather significant. 
The placement speaks to the relationship with the person. For an intimate loved one, the receiver might place its artifact proxy in close proximity and private location, such as a nightstand. For a colleague, the receiver might place its proxy at a location that mirrors their relationship, such as a desk.

Future systems could support this perspective by being aware of the relationships between the user and their friends. Current social media has developed relationships classifiers in the social network; for example, Facebook includes groups of friends based on closeness and how they met. This kind of relationship network can be extended to physically-grounded connections by mapping people's relationships to physical artifacts and spaces. With this knowledge, future systems could assist users maintain their connected physical environment meaningfully and orderly based on their personal relationship network.

Regarding wormholes acting as wearable connections, some people treating wormholes as proxies for other people expressed a desire for a wearable wormhole [P7b] or one they can carry around in their wallets so they could easily send and receive content from their partner instantly. As the connected artifact is kept on the user's person, its relative position with respect to environment is not significant. However,
%they do not care about its relative position with respect to their own environment, however, 
they might still be interested to keep track of how their partner's location changes throughout the day.

Future systems could design around the ability to get instant updates about their partner's location. For instance, if the receiver is using one wearable wormhole and the sender is using multiple location-based wormholes, then the sender's many endpoints can be used as a light-weight form of location tracker.
The receiver can be notified on their wearable wormhole whenever the partner moves to a different location. Compared to a GPS location tracker, this format gives the sender a great degree of control over what locations are being tracked and shared with others. 
%On the receiver's end, the notification can be displayed with multimodal and tangible media through a wearable wormhole. For example, if the sender is near the coffeemaker wormhole, the receiver may be notified by a pleasant scent of coffee.
% the course of the always-on connection. 

% We consider that this mediates the amount of importance people place on location as a factor. 

% systems for social connection can serve a means to reach someone (this mirrors the predominant form of connection via a phone, where oftentimes the position of the phone does not matter, just that the other person can be reached through it). 

% From the above analysis, we can infer that the physicality of connected artifacts are indeed valuable to the user who treated them as a proxy of person and future systems could draw inspirations from it . 

% The users, who prefers fast always-on connection, 

\subsubsection{\textbf{Artifacts to Create Overlapping or Connected Spaces}}
\label{Sec:OverlappingSpaces}

% Participants: 10b, 12a, 9a\&b

Some participants [P4a, P5a, P6b, P9a\&b, P10b, P12a] considered their connected artifacts to be a means of connecting their spaces with their partner's spaces, similar to the literal concept of a wormhole. By establishing wormholes with this viewpoint, they gained a sense of presence of their partner [P5a, P9a\&b, P10b] and appreciated that having connected artifacts could make it as though their friend can co-inhabit the space with them [P5a, P10b]. Some people also associate connected artifacts with the concept of leaving notes for each other around the house [P9a, P6a]. 
% However, P6a also described that having to receive transmission at specific locations creates a unique experience and a surprise feeling: 
% \begin{quote}
%     \textit{I felt closer to her and picture where she is throughout the day in her home.} - P5a
% \end{quote}
% \begin{quote}
% \textit{I like the idea of the desk marker because I'd be like, "Oh I'm at my desk and you're at my desk."} - P10b
% \end{quote}

% This can be likened to the experience of leaving or receiving notes for someone around the house [9a, 6b]
% % For instance, someone can leave a note about lunch on the fridge,
% in which the position of the note by the fridge is significant: 
% % \begin{quote}
% %     \textit{It'd be cool if [P9b] sent me a Bitmoji of himself with his coffee on my fridge or a small writing message that appears on the fridge as I'm going by the fridge.} - P9a
% % \end{quote}

% \begin{quote}
%     \textit{It reminds me of my parents would leave notes around the house...  I think it makes it feel more like a presence kind of thing. I'm in the person's house.
%     % when they were not there, where it would be like, "Oh, there's food in the fridge or whatever." You don't know, you're not expecting it and then you stumble on it. That aspect of it is really cool
%     } - 6b
% \end{quote}

For these users, being able to choose multiple artifacts that can be seamlessly integrated into different parts of their homes is helpful. The specific artifact that is chosen is less important than how the artifacts are positioned within the connected spaces. P4a and P9a commented on the desire to have a greater degree of spatial co-relationship, such that they can position transmitted objects in their partner's space mirroring positions in their own space. 

\begin{quote}
    \textit{[M]ake it a little bit more immersive and to feel like I'm in her space, I think it would have been cooler if it was more of a 3D image and if it was a specific place, especially if it could be wherever it was in relation to her wormhole. If she has her coffeemaker that's off to the right of her wormhole, so she takes that photo, I'd want to have to look at the same place to see where is her coffeemaker in relation to my wormhole.} - P4a
\end{quote}

% \begin{quote}
%     \textit{I could sent him the bouquet of roses from my wedding just placed on my counter-top, and then somehow transmitted to [P9b]'s counter-top.} - P9a
% \end{quote}

% This kind of perspective mirrors other works in the Ambient Media Space, focused on a sense of co-presence.

% \yy{learning about each other's environment}

% P9a\&b connected spaces like kitchen and laundry room, and transmitted objects from those connected spaces, such as drying sheets, blue apron meals, and snakes in the fridge. They are coworkers, have never been to each other's homes, by connecting spaces at home, they learned about each other daily life from another lens. They recall strong association between the objects they received and where they are being sent from.  

% \begin{quote}
%     \textit{``There was a level of presence like `Oh, he's been at the fridge or he's been there' (P9a). Yes. He's been here (P9b). I'm here, we're both here (P9a).''}
% \end{quote}

Symmetrical mapping of spaces is most intuitive and works best to support a sense of presence. Future systems could incorporate computer vision technology that plays a role in recognizing the similarity between physical spaces of both parties and makes suggestions to how spaces should be connected. Future systems could also use computer vision data to accurately map the spatial co-relationship between transmitted object and the connected artifact, enabling a 3D hologram representation of the transmitted object placed in the connected environment with meaningful spatial mirroring. 

Additionally, the number and placement of connected artifacts may have an effect on the sense of presence. Future systems may optimize by suggesting centrally located artifacts in a space as connection points, such that more surrounding objects can relate to them spatially. The number of connected artifacts could be scaled based on the system's understanding of the size of the user's environment and the spatial coverage of each connection point. An optimized number of connections would help achieve a symmetrical spatial coverage between connected spaces and therefore maximize sense of presence.

%However, in real life, often times paired users do not have the same spaces in the respective environment. 

%For example, if partner A has a garden while the partner B does not, the system might suggest that the optimal contextual symmetry mapping from Partner A's garden to Partner B being a particular window sill where their houseplants are placed.
%The layout of physical spaces are often dissimilar between partners even if there are the same type of spaces, making it challenging 

% Features 
% Number should map to the physical environment 
% Mapping - representative objects to representative objects (Contextually similar), Best to have symmetrical mappings 
% E.g. garden to window sill 
% Hyper-awareness of space (and relative spatiality) 
% Having more wormholes in a space might help to increase the feeling of spaces overlapping or being shared 
% Coverage range of wormhole / Accessible Location for the wormhole can change how ‘connected’ spaces feel to others?
% Coverage can scale with number, or location centrality 
% System should suggest centrally located artifacts to serve as connection points, such that they have a more perceivable relative position to other objects in the space. 
% System should have understanding of surrounding spatial environment? 
% Future connected objects have an understanding of where they are in relation to other things (could be Lidar, if the object is smart) or if it’s a glasses system, it also builds and understanding of the spatial layout) 

\subsubsection{\textbf{Artifacts for Sharing an Activity}}
\label{Sec:SharingActivity}
% Participants: 10a, 10b, 12a, 9b, 5a&b, 7b

% Many prior works on remote social connection have tackled the idea of being able to bond over a shared activity over distance. Connecting over cooking or dining is a particularly popular theme \cite{barden2012telematicdinner,chai2017cookingdistance, nawahdah2013virtuallydiining}. 
Several participants [P4a, P5a\&b, P7a\&b, P9b, P10a\&b, P12a] considered their connected artifacts to be places in which to participate in a shared activity with their partner, aligning with many prior works on sharing activities over distance~\cite{barden2012telematicdinner,chai2017cookingdistance, nawahdah2013virtuallydiining}, in which cooking and dining are popular themes.
% for which to share were also keen on the idea of being able to connect through the same activity as their friend. 
By employing symmetrical connections, these participants vividly perceived themselves sharing activities such as cooking [P5b] or working [P12a] together. Participants also dedicated wormholes for the sole purpose of sharing special hobbies (piano [P4a, P10a]) and events (pool party and baseball game [P7a\&b]).
%In case special events away from their usual wormhole locations, participants wished to have a portable or temporary connection such that they would not miss out sharing the interesting activities[P7a\&b].

%moving these to Results. 
% P12a enjoyed connecting with their partner while working: 
% % over food as well as other daily activities such as working. P12a commented on 
% \begin{quote}
%   \textit{ I can imagine himself typing when I'm typing when he's sending me different types of stuff from the desk}. - P12a
% \end{quote}

% Through a sequence of asynchronous transmitted artifacts, some users perceived a continuous line of daily activities from their partner:
% \begin{quote}
%     \textit{``I saw Sparkles coming from their door and their coffee machine, so I figured she must be starting her day; I saw lots of Sparkles by the kitchen and door, which made me think she was cooking before she left. ''} -- P4a
% \end{quote}

Many participants enjoyed connecting asynchronously through activities. They interpreted a stream of transmissions from their partner's different wormholes as continuous daily activities [P4a], which gave them a strong sense of co-presence with their partner.
% P9b primarily belongs to a connecting spaces pattern, however, they commented on the desire to experience physically grounded storytelling by receiving messages from multiple wormholes in an order that mirrors the sender's sequence of daily life. They also imagined that transmitting several objects related to an activity could help them communicate what they were doing.
Building on this, P9b desired to have sequencing and grouping features to better illustrate how their partner transitioned between activities.

\begin{quote}
    \textit{"I would like to see it as a story of five different ghosts, in order of which they were transmitted back to back. From my perception, that would give me a sense of seeing their life together [...] If you tell the system what you are doing, then it can group those objects by story. If I'm cooking then I want send those multiple things from my kitchen, groups them together. "}- P9b
\end{quote}

Interestingly, our technology probe was designed to support asynchronous connections, and yet, some users gravitated to having synchronous experiences, underlining the significance of supporting synchronous activities while maintaining the poetic sense of asynchronous ambient connections.

%\joanne{Do we want to make it about syncing over an activity, or that they can read into the artifacts about activities?}

\begin{quote}
     \textit{"...suddenly I knew we were both live in the moment together, that would have been maybe meaningful where we could connect, but there was really no way to align. Are we two ships sailing passing in the night or are we having a moment because we're here in the moment together?"} - P10a
\end{quote}

Future systems for supporting the notion of shared, connected activities should identify which common activities both connected partners engage in and suggest a symmetrical mapping of connection points in the vicinity of that activity to the users. 
% As we learned that sense of connection is stronger if the connected artifact affiliates the activity directly. 
Future systems could identify activity-signifying artifacts (e.g.piano) as its suggestions. 
There is also opportunity to recognize the context of the activity in order to determine whether users would desire a synchronous or asynchronous sharing experience. 
% During occasional special events, the system could also automatically support the creation of a temporary wormhole if the events are away from existing wormholes. 
% There are times when people may want to share about special events.
% For example, P7a wanted to share special moments at a pool party and brought the wormhole with them to the pool. In this case, it could be helpful if the system supports the creation of a temporary wormhole. 
% The number of connection points usually mirrors the number of activities a user engages in, but non-routine event might require activating a temporary portable connection point to facilitate sharing that activity. 
% Some activities may desire the system to be scalable in order to support
Last, some activities may benefit from one-to-many broadcasting or many-to-many connections. A multi-location alumni reunion is a good example of such an activity. Future systems could explore how to support such synchronous and scalable connections, and how to mitigate multiple sequences of activities in a network of physically-grounded connections.

% P5a&b specifically interpreted Sparkles as they each "looked" at the wormholes. 

% P5b transmitted a highlighter as it signifies the activity of note-taking. 
% P7b interpreted kitchen location indicator as their partners is engaged in cooking.

% \joanne{sending highlighter, because she's taking notes 5}
% \joanne{playing with cat}
% \joanne{she looked at it [wormhole] as well...}
% \joanne{I did see a lot of sparkles, so I know that she looked at it often...}

% \subsubsection{Peek-a-Boo Model}
% Where you can peer into someone else's world.

% \subsection{Overall Remarks}
% \subsection{Mixture of Perspectives}
%\subsection{Blended Patterns of Behavior}
\subsection{Mosaic of Behavior Patterns and What Connected Artifacts Represent}
\label{sec:mosaicofbehavior}

%In the context of ubiquitous computing for social connections, 
Emerging naturally from using the customizable technology probe, 
our findings reveal that people adopt multiple behavioral patterns and find values in all of them simultaneously.

Some participants predominantly displayed a ritualistic behavior pattern, but also enjoyed rare moments of serendipitous playfulness. P5b, as an example, found value in learning routine details about their partners daily life, but also took a spur-of-the-moment experiment using their cat as a connected wormhole. In some cases, the two partners in participant pairs adopted opposing communication patterns while connected in the same ubiquitous ecosystem. For example, P10a is serendipitous, while their partner P10b is ritualistic.

Participants also adopted different viewpoints for what each of their connected artifacts represent. For example, P7a established some connected artifacts acting as proxies of their partner and other connected artifacts acting as places from which to share an activity. Their favorite beverage served as a proxy of the partner on a daily basis, but during an interesting event such as pool party, they dedicated a wormhole to being a way to transmitting activities taking place during the event.

Given this behavior, we think that there is not one superior behavior pattern that physically grounded social connection technologies should cater to. Rather, we anticipate that many of these different perspectives can co-exist in people's lives and in ubiquitous computing as a mosaic of different types of connected artifacts.

\subsection{Other Social Considerations}
\label{sec:socialconsiderations}
% \joanne{Still under construction...}

%\yy{privacy, loneliness, overwhelming attention/ surveillance culture. }

While we focused on patterns of behavior that people exhibited, users also brought up considerations that are important to the design of future ubiquitous computing systems for social computing. 

\subsubsection{\textbf{Privacy}} %This is written from a  sender's perspective
\label{sec:privacy}
Generally, people were comfortable sharing with one another since they were friends, but comments  regarding privacy did arise [P6a, P9b, P10 a\&b]. Some users hesitated or had questions about what the glasses captured and what we as researchers would be able to see. Future systems will likely need to clarify how privacy is maintained and protected by companies providing such services. We also anticipate that as the number of connected artifacts scales upwards, people may lose oversight of what objects are connected and sharing information. Furthermore, while our technology probe used printed markers, it's possible that future systems can invisibly tag artifacts. This can exacerbate this potential issue. One possible solution for future investigation is establishing tools or methods for automatically curating and reviewing the information that is being shared through a person's network of connected artifacts. Tools that can give users an overview of all of their connections may be particularly useful. 
\subsubsection{\textbf{Overwhelming v.s. Loneliness}} 
% This is written from a receiver's perspective
% Given the attention economy, we also recognize the importance of not adding to distraction. 
\label{Sec:OverwhelmLoneliness}

Despite there being more platforms than ever for online social connection, there is nevertheless a growing epidemic of loneliness. Furthermore, there is a problem around \revision{having too many communication channels demanding our attention and effort}. These negative effects could extend into the realm of physically-grounded and connected artifacts, as one participant had speculated:

% It is also important to consider possible negative consequences of a landscape of physically-grounded and connected artifacts:

\begin{quote}
    \textit{I feel like too many may feel a little overwhelming and intrusive. Then it turns into more of an inconvenient scavenger hunt, because yes, you constantly have to go around and look at a zillion different little things that you scan is or like, it could give you a bad reaction of loneliness, if you don't receive anything and you have a zillion different wormholes and you don't get anything. I feel like keeping it semi-small is maybe better.} - P9b
\end{quote}

% One flexibility we explored with the technology probe is the number of connections. As P9b suggested, the number of connections could be correlated to the activeness of one's social network. 
To address this issue, future systems might play a more active role in curating the number of available connections. They may be able to make suggestions to users and/or automatically increase or prune unused points of connection to optimize in accordance with users' needs and usage levels.
\revision{Future systems might position themselves as streamlining and unifying communication rather than adding even more channels. As an example, it could be possible to surface a photo or video message from a friend sent on any communication channel (text, email, Slack, etc.) as a Ghost in a relevant location, then automatically forward any Sparkle or Ghost sent in response to whichever channel the original message came from.}
% \joanne{In addition to the possible overwhelm that could be  incurred when using the system, such feelings may be compounded when multiple channels of communication are active simultaneously (e.g. phones). It may be necessary for future systems to include mechanisms to manage information flow to reduce duplicates and mitigate overwhelm.}

% to strike an optimal setup that meets users' needs and usage levels. 

% More active social network might demand larger number of wormholes; less active social network might need only a small number of wormholes. The frequency of messages received from each wormhole should be used by the system as means to optimize the number of connection, in order to strike a balance between loneliness and overwhelm.
% what could the optimization look like
% suggest additions? automatically prune unused ones? filter out old messages to improve relevancy?
% (these two ends of this spectrum?)

% In the case of this study, none of the participants complained about being distracting since they still only wore the glasses for a fraction of each day, but this may need to be considered if platforms are constantly within reach. 

\section{Discussion}

\revision{
    Our study had several key takeaways, including
    % that Social Wormholes added to participants' feelings of social connection and presence (Section 5.1), 
    % that participants used a variety of household objects and locations as endpoints motivated by different reasons (Section 5.2), 
    % that they connected objects in both symmetric and asymmetric fashions and had different ways of interpreting their connections (Section 5.3), 
    % and that they tended to use a few wormholes each day (Section 5.4). 
    % We also learned 
    that users engage in several different behavior patterns for connecting objects and communicating through them (Sections 5.2 and 6.1), that users assign different roles to physical artifacts (Sections 5.3 and 6.2), and that each user may adopt these varied patterns simultaneously (Section 6.3). Table \ref{tabledesignimplications} expands on these findings in more detail.
}
% \joanne{New Discussion Intro}
% Intro to Discussion
% \joanne{(1 - ADDRESSED )We approach this problem from the user side, not designer side, finding users naturally gravitate towards the choices. % (2 - ADDRESSED )And, ecosystem perspective, studying everything in cohesion, finding a mosaic of perspectives.}
% (3 - ADDRESSED) - Not a single pair but an ecosystem of them

% (6.3 Mosaic)
% \joanne{Our results shed light on (number) revealed six new insights on ubiquitous social awareness that prior work didn't explore. Ex. Fourth, to our knowledge our study was the first to compare... prior work, blah...}

% 5.1
%Generally, we observed that our system, like many other systems have improved self-reports of social-connection between remote friends \cite{---}, and a sense of presence \cite{---}.
% Prior works that study ambient social systems found that their users experienced an improved sense of social connection \cite{?} and sense of presence \cite{?}. Our technology probe Social Wormholes, although did not limit its usage to ambient display, has proven to support the same improvement in their users.

%  \cite{li2018reviewemotionalcommunicationLDR}
% \cite{chang2001lumitouch, siio2002peekadrawer, goodnightlampdesign}
Our investigative approach was unique from prior works in two ways. First, the vast majority of prior studies on the use of tangible or ambient systems for social connection have comprised a single pair or set of near-identical connected devices between remote parties \cite{siio2002peekadrawer, dey2006awarenesstoconnectedness, chang2001lumitouch, kim2015breathingframe}. In contrast, by creating and deploying \textit{Social Wormholes}, we gained knowledge into how ubiquitous computing---namely, an ecosystem of multiple artifacts--- can be used. As expected, we found that ecosystems of multiple endpoints improve social connection and a sense of presence (see Section \ref{SectionGeneralObservations}), similar to the range of single-connection systems. However, by studying the use of multiple lines of connection in concert with one another, we were in addition able to outline and distinguish between different patterns for social connection, as well as gain novel insight into how these patterns can co-exist with one another.
Second, prior social connection systems have been designed based on researchers' own design intuitions or through co-design processes with users \cite{dey2006awarenesstoconnectedness}. In other words, prior system configurations have largely been determined by researchers. \textit{Social Wormholes}, instead, empowered pairs of friends to design and curate their own set of existing objects as endpoints for social connection on their own. This allowed us to learn what other objects and configurations may be fruitful, beyond those that have been previously identified, based on the usage patterns that emerged organically over the course of our two-week field study.

Stemming from our results, we discuss and contribute five points of knowledge to the prior literature
%provide a set of five new insights 
regarding ubiquitous social connection systems. %designate and maintain

First, we discuss that users value choosing their own artifacts. 
Designers may not always choose the artifacts that users themselves would and users would value this autonomy. With \textit{Social Wormholes} enabling this ability, we found that several of the objects and places participants chose overlapped with those from prior literature (e.g. kitchen \cite{chai2017cookingdistance}, bedroom \cite{siio2002peekadrawer}, and mirrors \cite{schmeer2010touchtracemirror, dey2006awarenesstoconnectedness}) (see Section \ref{sec:results-what-artifacts}), thereby reinforcing that these are desirable to be used as endpoints for social connection. Furthermore, while our participants did not choose objects such as picture frames \cite{chang2001lumitouch, kim2015breathingframe}, some of their comments underlined the appeal of using personal tokens that are specially representative of their remote friend (see Section \ref{SectionProxiesforPeople}). On the other hand, there were artifacts that our participants chose to connect that have not been used before in prior systems, such as toilets for humor and pets. These choices corresponded more with the desire to share fresh content (see Section \ref{SectionSerendipitous}). Ultimately, more study data is needed to capture the full extent of artifacts that would or would not be ideal for social connection. Nevertheless, future ubiquitous systems should therefore give users more autonomy to choose and update their own artifacts, not only to account for user's unique preferences and circumstances but also to support changing trends for artifacts, since users' ideas of what artifacts are interesting or fresh can change over time.

Second, we discuss that the surrounding context of routines and other people in the space shapes users' choice of artifacts. 
Routine was an important factor for users when choosing their artifacts (see Section~\ref{SectionHowDidPeopleChooseArtifacts}). This strongly aligns with Cao et al.'s \cite{Cao2010} highlighting of routine as a theme in remote communication. However, while some users allowed transmissions to occur as an accompaniment to their daily activities (see Section \ref{SectionRitualistic}), others made it a deliberate and standalone practice to send transmissions  (see Section \ref{SectionSerendipitous}). We also discovered that users are sensitive to how their choice of artifacts may affect or be perceived by other members of their household. To the best of our knowledge, our findings in this area are novel since most previous studies have focused on single users \cite{grivas2006digitalselves, hakkila2018connectedcandles, heshmat2020familystories, kim2015breathingframe, siio2002peekadrawer} or family units \cite{Judge2011, Judge2010} rather than studying the system in the context of a mixture of both users and non-users in the same household. Future designers should therefore anticipate that routines and other people, both users and non-users, are contextual elements that will shape users' choices and behaviors.

Third, we discuss the value of spatial context based on users' perspectives. Spatial information is most valuable to people who seek to connect through overlapping spaces (see Section \ref{Sec:OverlappingSpaces}) or through shared activities (see Section \ref{Sec:SharingActivity}).  %A similar spatial context with a remote friend forms an intuitive foundation for creating a sense of presence. To amplify this effect, 
Participants expressed a desire for transmissions to be presented in true spatial relationship to endpoints, expanding the perceived footprint of shared space and making transmissions more real and immersive. This echoes prior works that stress the importance of spatial coherency \cite{kim2017effectsofVHspatialandbehavioralcoherence} and the preservation of physical-space information in virtually-mediated exchanges \cite{fender2022causalitypreservingAR}. When connecting over a shared activity, the spatial information is valuable for adding a greater feeling of commonality and relation during an activity. For instance, participants enjoyed the illusion of their friend typing at their desk just as they were typing at their own desk. The spatial information adds an extra layer of social awareness \cite{sereno2020collaborativeworkinAR, hassenzahl2012all}. In contrast, when users center their perspective on the artifacts as proxies for people(see Section \ref{Sec:ProxyforPeople}), they value the space as a context to the closeness of their relationship, and additionally, they value timing context of the messages. Thus, future designs should cater to capturing and relaying spatial information to enhance the user experience based on their perspectives.

Fourth, our findings add to prior literature on the preference around asymmetric spatial connections \cite{Asymmetry2008}, in addition to symmetric connections. In contrast to the common assumption that people would prefer symmetrical pairings, a surprising 72.2\% of connections with \textit{Social Wormholes} were asymmetrical (see Section \ref{SectionHowAreArtifactsConnected}). Prior works on tangible systems for social connection have often been designed in symmetric pairs \cite{li2018reviewemotionalcommunicationLDR}, placing heavy assumptions that symmetric connections are preferred. Our investigation directly studies how users choose to connect artifacts to one another and reveals the comparison between symmetric and asymmetric connections.

%In the case of connecting over a shared activity, we find that symmetric spatial context can help achieve a shared sense of space and presence. However, 
We find asymmetric connections offer three advantages. First, asymmetric connections do not require effort to coordinate artifact pairs between casual friends. Depending on the closeness of their relationships, close friends and romantic partners may associate their relationships with a spatial context (see Section \ref{Sec:ProxyforPeople}) and therefore choose to coordinate symmetric connections. Second, asymmetric connections can afford a serendipitous and fun experience, when users receive surprising messages at unexpected locations (see Section \ref{SectionSerendipitous}). Third, asymmetric connections can broadly support the mosaic of perspectives, as we find that the two partners in a connection pair often adopted opposing communication patterns (see Section \ref{sec:mosaicofbehavior}).

Lastly, we discuss that distributed endpoints encourage opportunistic sharing and ambient receiving. 
%had different effects on the sending and receiving experience. 
From the sending perspective, distributed endpoints provide greater coverage around users' space and daily routine, making it easier to immediately share moments about their lives. 
%From the sending perspective, many users established endpoints that complemented their daily routine, making it easier to immediately share details about their everyday lives. 
% The multiplicity of endpoints in 
As such, \textit{Social Wormholes} worked well to encourage sharing of more routine or ``mundane'' activities \cite{tolmie2002unremarkablecomputing} during ``empty moments'' \cite{lottridge2009sharingemptymoments} in their everyday lives. 
From the receiving perspective, the distribution of endpoints enabled people to receive messages when their time and place were relevant, which catered well to people who enjoy ambient receiving. 
These users focus on going about their daily life and perceive social communication as an accompaniment in the background space \cite{mankoff2003heuristic}.  
%enjoyed connecting over shared activities (see Section \ref{Sec:SharingActivity}). 
%Depending on the users' individual perspectives, this however also came with a cost. 
Depending on the users' individual perspectives, the wide spatial distribution however also compromises the desire for instancy. Having to receive the message at the designated location introduced a potential time delay and required users to physically arrive at the corresponding endpoint to receive transmissions. In turn, it made some users feel that it was more difficult to ``keep up'' with all of the information (see Section \ref{SectionRitualistic}), potentially leading to feelings of overwhelm (see Section \ref{Sec:OverwhelmLoneliness}). Future designers should consider when it might be helpful to limit the number of connections and/or incorporate notification features into a ubiquitous social connection system to strike a balance between ease of use for sharing and feelings of control in receiving information.

\section{Limitations and Future Work}
% keep this tight, or expand to more grandiose context?
%\joanne{Speculation of a system that is more proactive in making suggestions, or adding spontaneous Wormholes}
%\joanne{Covid and doing the study restricted to the home since everyone was homebound...}

Our study participants generally followed a work-from-home routine due to the COVID-19 pandemic, so our results apply to connection endpoints being used in this context. \revision{This unusual circumstance may have biased the participants' towards more routine patterns of behavior, since people were encouraged to stay at home.} In the future, it would be interesting to investigate ubiquitous connection endpoints in other contexts. Experimenting within professional office settings or public outdoor events, for instance, could yield additional insights into unique behavior patterns.
% Given this situation, they may have spent more time at home than they normally would, which may have influenced how frequently they used the system throughout the day.
% Limited diversity / less 'novelty' in items sent as ghosts?
% Limited the focus of our investigation to domestic life?
% Emotional influence? COVID might have made people more sad --> this would lead to what kind of behavior though?
%daily routine
 % Limited study environment to homes 
 % Possibly would have biased people on "routine" behaviors
% not sole member in household...for example, some participants were placing markers in private spaces in consideration of their household members
%future investigation to study group social behaviors.

% \revision{We recruited our participants from employees of a technological company due to the limited access to the AR glasses hardware and the ease of distribution for field study. Although our participants come from diverse departments with technical and non-technical roles, there could potentially be a limitation in that our participants are generally more interested in technology than the general public.}
\revision{We recruited our participants from within a technology company, since the AR glasses hardware at the time of this study could not be distributed to the general public. Although we recruited participants with both technical and non-technical roles in the company, they were likely more interested in technology than average, and may therefore have been more biased in favor of the experience overall.}

Our research is limited in that participants were suggested to use the AR glasses for a minimum of 15 minutes per day. This was to help ensure that each person would be able to try out the technology long enough to understand the features and see how it might be useful or create friction between them and their remote partner. We also wished to uncover knowledge about how usage may change over a duration of time, so we requested participants to use the technology each day throughout the two-week field study duration. As this may have influenced our findings, it would be valuable to revisit this topic in the future when wearing AR glasses regularly becomes more commonplace. Due to the COVID-19 pandemic and the widespread locations of participants, we were also unable to conduct real-world observations of participants using the technology probe that could have added richness to our findings.

% \joanne{Check here again.} \revision{As part of our study procedure, participants were asked to use the glasses for at least 15 minutes per day. They could determine whether they would wear it in short bursts, or in one longer session. Since AR glasses are still not widely adopted and regularly used, they are not yet part of people's daily habits. Therefore this request was necessary to conduct an early exploration about future ubiquitous systems, which of course may have influenced study results. It w} 

% \joanne{As part of our study procedure, we asked participants to use the glasses at least for 15 minutes per day. Although we were interested in understanding how this system would be used daily, AR glasses are not yet broadly adopted by people to be worn all day. Therefore, this request was necessary to begin to ask this question.}

% \revision{In our study, we restricted artifacts to only be connected in a one-to-one fashion. However, it would be interesting to broaden the design space to one-to-many or many-to-many connections between augmented physical artifacts. Additionally, we focused our technology probe to work only between two friends, although people naturally interact with a larger circle of friends. Hence, there is an opportunity for future work to investigate how an ecosystem of connected physical artifacts can be used to nurture a person's social connection to a larger network of friends.}
In our study, we restricted artifacts to only be connected in a one-to-one fashion. While we used one-to-one connections as an initial starting point, it would be interesting to explore one-to-many or many-to-many connections between augmented physical artifacts, as mentioned in our design implications - particularly for serendipitous communication. 
% Additionally, we focused our technology probe to work only between two friends, although people naturally interact with a larger circle of friends. 
Hence, there is an opportunity for future work to investigate how an ecosystem of connected physical artifacts can be used to nurture a person's social connection to a larger network of friends. 

\revision{Additionally, our study focused on one-to-one communication between two friends, primarily due to a limited number of hardware units available to support our study. Nevertheless, some participants brought up how sharing their living space with others influenced how they used \textit{Social Wormholes}.
% People often share living spaces with family members. our findings suggested some participants' behaviors around using Social Wormholes were influenced by the presence of other people in their homes. 
For example, some participants mentioned that they were mindful of others in their households when deciding where to place their markers. 
% hiding markers in private spaces only.
%However, people often share living spaces with others and interact with a larger circle of friends and family. 
% Indeed, communicative technologies play a pivotal role in facilitating people's support networks \cite{beckman_dreams_2020}, which is essential in an age where people must play multiple dynamic roles.
Furthermore, we live in an era in which people do not only share their home environments, but also, adding to the complexity, face the demand of juggling multiple roles, such as being a parent, a friend and a colleague. In this case, communication technologies play a pivotal role in helping people maintain a network with others as supportive scaffolding in their dynamic home environments \cite{beckman_dreams_2020}. We suspect that the behavior patterns and perspectives we observed in this study may extend to these more complex communication scenarios -- in particular, the tendency for people to engage in a mosaic of these. Exploring these themes in the context of negotiating over shared spaces and managing the division of roles would be an interesting line of exploration for future research.}

We also see the opportunity for existing social applications to extend their reach beyond the digital domain and adopt an aspect of being physically grounded. Social networks could explore how synchronous and asynchronous updates about friends' lives can be shown through physical artifacts in someone's environment. For example, while existing gaming platforms often notify users when their friends start playing a game, such information could be anchored to a physical gaming console. Notifications could also be displayed with multimodal and tangible media. If a sender is near a coffeemaker wormhole, for example, the receiver might be notified by a pleasant scent of coffee. Furthermore, the present study focused on supporting social connections between friends, but it is also possible for updates from non-human entities (e.g. pets, companies, world news) to have a physical footprint in the home via associations with different artifacts.

%\joanne{Comment about how to cater to a mosaic of needs.}

%There is opportunity for building a future ecosystem in this design space, where a mosaic of designs and products that bring value to individual behavior pattern can amplify every day values collectively.

%Future designers could acknowledge this aspect by designing flexibility into artifacts to cater to these different usage patterns. Portability?

% New features would have to be built to bridge the digital world with the physical one. 

% Notification 

% Other types of apps can explore physically-groundedness, configurable-ness 
% Other social connection apps might want to use our insights to make themselves more physically grounded
% THis type of investigation that we did could be used to look at the design of notifications (anything that pops up on notifications center on iphone)
% Notifications that are not just human?
% Btw - steam notifies what game your friend is playing

\section{Conclusion}

As the potential for computing to blend into our everyday environments grows, powerful opportunities open up for social connection. Inspired by the vision of ubiquitous computing, this paper explored how distributed and physically-grounded social connections can be used to help family and friends stay connected. We implemented \textit{Social Wormholes}, an AR-based technology probe that enabled pairs of friends to craft an ecosystem of connected physical artifacts through which to transmit information. The probe allowed users to choose what artifacts became endpoints for connection, how they were mapped to one another, and how many endpoints they wanted at any given time. Based on logs, surveys and interview data collected in a field study with 24 participants, we uncovered different perspectives and patterns of behavior for engaging with such a system. This includes ritualistic versus serendipitous behavior patterns as well as differing perspectives around whether these physical artifacts serve as a proxies for friends, as overlapping spaces, or as opportunities to connect over activities. Furthermore, we found that while different patterns of behaviors exist, users often adopt a personalized mosaic of these patterns. Our findings point to a future of ubiquitous, physically grounded social connection that is not one-size-fits-all but rather individualized to each person's values.
\bibliographystyle{ACM-Reference-Format}
\bibliography{refs}

\end{document}